\documentclass[preprint2]{aastex}

\begin{document}
\title{Dust Evolution in the Dwarf Galaxy Holmberg~II}

\author{D.S. Wiebe \altaffilmark{1,*}, M.S. Khramtsova \altaffilmark{1}, O.V. Egorov \altaffilmark{2}, T.A. Lozinskaya \altaffilmark{2}}

\altaffiltext{1}{Institute of Astronomy, Russian
Academy of Sciences, Pyatnitskaya ul. 48, Moscow, 109017 Russia}

\altaffiltext{2}{Lomonosov Moscow State University, \protect\\
Sternberg Astronomical Institute, Universitetskii pr.~13, Moscow, 119992
Russia}

\altaffiltext{*}{E-mail: dwiebe@inasan.ru}

\begin{abstract}
A detailed photometric study of star-forming regions (SFRs) in the galaxy
Holmberg~II has been carried out using archival observational data from the
far infrared to ultraviolet obtained with the GALEX, Spitzer, and Herschel
telescopes. Spectroscopic observations with the 6-m telescope of Special
Astrophysical Observatory of the Russian Academy of Sciences are used to
estimate ages and metallicities of SFRs. For the first time, the ages of SFRs have been related to their emission parameters in a wide spectral
range and with the physical parameters determined by fitting the observed
spectra. It is shown that fluxes at~8 and 24~$\mu$m characterizing the
emission of polycyclic aromatic hydrocarbons (PAHs) and hot dust grains
decrease with age, but their ratio increases. This implies that the relative
PAH contribution to the total infrared flux increases with age. It is
suggested that the detected increase in the ratio of the fluxes at~8 and
24~$\mu$m is related to the growth in the PAH mass due to destruction of larger grains.

\end{abstract}


\maketitle

\section*{INTRODUCTION}

Despite its relatively small mass fraction \mbox{($\sim1\%$)}, dust plays an important
role in the evolution of galaxies both on a global scale and in individual
star-forming regions (SFRs). It is also a valuable diagnostic tool, because in late-type galaxies a
significant fraction of the bolometric luminosity is
generated by dust grains of various types.

Among the various interstellar dust components, the so-called polycyclic aromatic
hydrocarbon (PAH) macromolecules, which are presumably responsible for the
formation of unidentified infrared (IR) bands (UIBs), have attracted
particular attention in recent years. Their IR emission is believed to be
generated by the reradiation of absorbed ultraviolet (UV) photons (Draine and
Li~2007) and, therefore, can serve as a star formation rate indicator
(Calzetti~2011). However, the utilization of PAHs for this purpose is hampered by
the lack of necessary understanding of their formation and
destruction processes. The key problem is the twofold role of the UV
radiation in the evolution of PAHs. UV photons excite IR~transitions in
PAH molecules but also simultaneously destroy these molecules. Therefore, the
dependence of the PAH band intensity on UV~radiation can be nonmonotonic.

One of the best-known properties of PAHs is the dependence of their relative
abundance on metallicity. Specifically, the $q_{\textrm{PAH}}$ parameter
characterizing the PAH fraction in the total dust mass decreases from a few
percent in solar-metallicity galaxies to a fraction of a percent in galaxies
with an oxygen abundance $12+\log({\textrm{O/H}})$ lower than~8.1 (Draine
et~al.~2007; Madden et~al.~2006; Hunt et~al.~2010; Engelbracht et~al.~2005). It
was shown in several studies (Gordon et~al.~2008; Wiebe et~al.~2011; Khramtsova
et~al.~2013) that this correlation not only is observed for galaxies as a whole
but is also traceable for individual SFRs inside galaxies.

A detailed study of the IR~emission in individual extragalactic SFRs has
become possible owing to the Spitzer and Herschel space observatories. The
angular resolution of their instruments is sufficient to investigate the
spatial distribution of emission from dust grains of various types in
star-forming complexes (though insufficient to investigate individual
HII~regions). Detailed mapping of the IR~emission, the dust temperature, and
the PAH mass fraction ($q_{\textrm{PAH}}$) allows relating these quantities to other {\emph{local\/}} properties of SFRs, for example, to the
intensity of their UV~radiation and metallicity. However, despite the large volume
of observational data, low-metallicity galaxies have not yet been adequately
studied in this respect.

Based on Spitzer and Herschel observational data and published metallicity
data, Khramtsova et~al.~(2013) investigated the relation between the oxygen
abundance $12+\log({\textrm{O/H}})$ and the PAH mass fraction
$q_{\textrm{PAH}}$ in more than 200 extragalactic SFRs. These authors showed
the PAH abundance in high-metallicity SFRs to increase with oxygen abundance
and concluded that the PAH abundance--metallicity correlation is most likely
related to their more efficient destruction in the interstellar medium with lower abundance of heavy elements. However, the data turned out to be insufficient to analyze in detail the behavior of $q_{\textrm{PAH}}$ in the lowest-metallicity SFRs.

This paper presents a detailed study of the dwarf galaxy Holmberg~II
(Ho~II). As this galaxy was included in several large surveys, for example,
SINGS (Kennicutt et~al.~2003), THINGS (Walter et~al.~2008), HERACLES (Leroy
et~al.~2009), and KINGFISH (Kennicutt et~al.~2011), it was thoroughly mapped in
many spectral ranges, from X~rays to 21~cm. Both Spitzer
observational data at~8 and 24~$\mu$m and longer-wavelength Herschel data,
which allow characterizing not only the PAH abundance but also the abundance of larger dust
grains are available for the galaxy.

Numerous ``holes'' and expanding supergiant shells (SGSs) are a
characteristic feature of the HI distribution in Ho~II. Puche et~al.~(1992)
isolated 51 SGSs in the galaxy with a total kinetic energy of
about $10^{53}$~erg. A slightly smaller number of shells (39) were identified by
Bagetakos et~al.~(2011) using the different criteria to classify a structure as a
hole or a supershell; they mostly coincide with SGSs from Puche
et~al.~(1992). The shell sizes vary between 0.26 and 2.11~kpc, corresponding to
a kinematic age from~10 to 150~Myr for expansion velocities of 7--20~km~s$^{-1}$
(Bagetakos et~al.~2011).

It was previously thought that the origin of a supershell is related to the combined action of stellar winds and supernova explosions in a young stellar association. However, the search for parent associations in Ho~II shells undertaken by Weisz et~al.~(2009) using detailed multicolor photometry based on Hubble Space Telescope observations was unsuccessful. More precisely, several stellar groups of different ages were found within many of the shells. Weisz
et~al.~(2009) concluded that the giant cavities in the HI distribution were
produced not by one star cluster but by several generations of stars that had
been born over hundreds of millions of years.

Thus, the structure of the interstellar medium in Ho~II indicates that active
star formation has taken place in the galaxy over the last several hundred
million years. Hodge et~al.~(1994) revealed 82 HII~complexes in the galaxy
concentrate mainly in the directions of the highest HI column density and
form several chains of active SFRs, the brightest of which are located
in the eastern chain. Another study of HII~regions in the galaxy Ho~II was
undertaken by Stewart et~al.~(2000). These authors estimated the ages of 45
HII~regions and divided them into four age groups. Most of the regions fell
into the groups with an age younger than 6.3~Myr. This is consistent with other
estimates of the star formation history in Ho~II, according to which the star
formation rate in the galaxy has increased over the last several million years
(Weisz et~al.~2008). This implies that we have an opportunity to study in
detail the evolution of young SFRs in a low-metallicity galaxy.

Here, we investigate the dust component of Ho~II and relate its properties
to the ionized-gas metallicity, the ages of SFRs, and the emission parameters
in them. We use GALEX, Spitzer, and Herschel archival observational data
for Ho~II and the observations of Ho~II with the Russian 6-m telescope at the
Special Astrophysical Observatory of the Russian Academy of Sciences (SAO RAS).

\begin{figure*}[t!]
%%% Figure:1
\center{~~~~~~~~~(a)~~~~~~~~~~~~~~~~~~~~~~~~~~~~~~~~~~~~~~~~~~~~~~~~~~~~~~~~~~~~~~~~~~~~~(b)}
\includegraphics[width=0.5\linewidth]{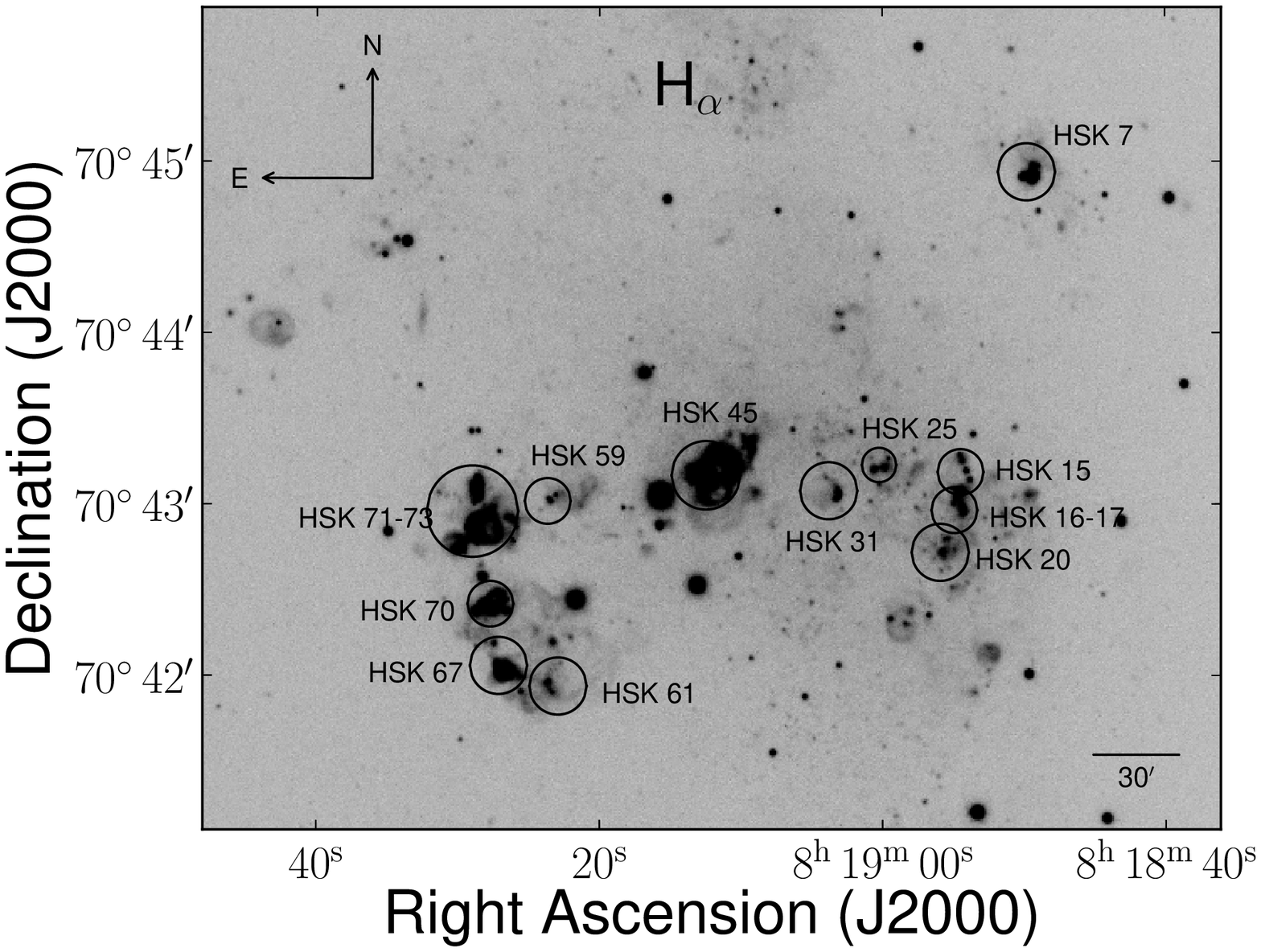}~\includegraphics[width=0.5\linewidth]{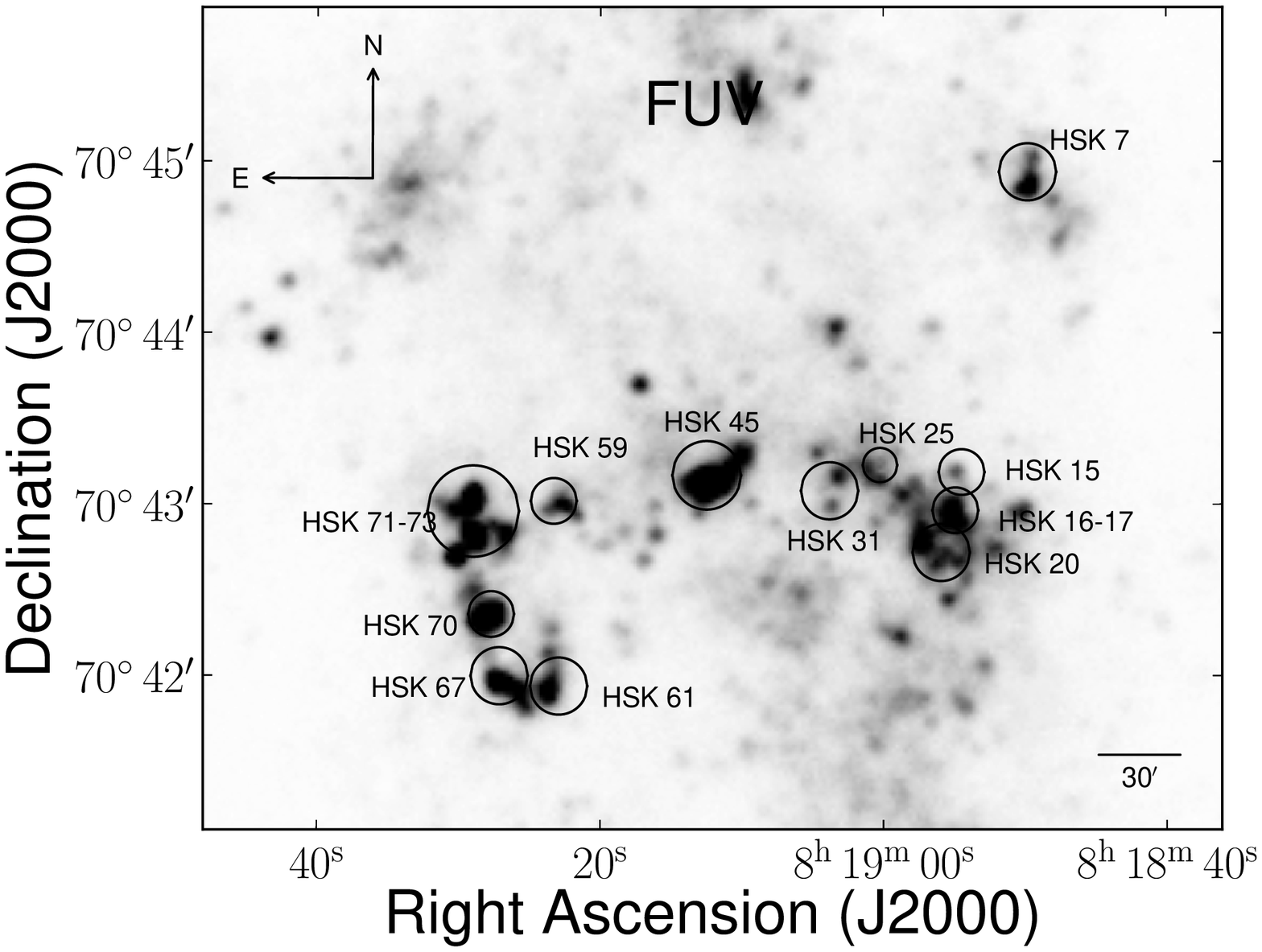}

\center{~~~~~~~~~(c)~~~~~~~~~~~~~~~~~~~~~~~~~~~~~~~~~~~~~~~~~~~~~~~~~~~~~~~~~~~~~~~~~~~~~(d)}
\includegraphics[width=0.5\linewidth]{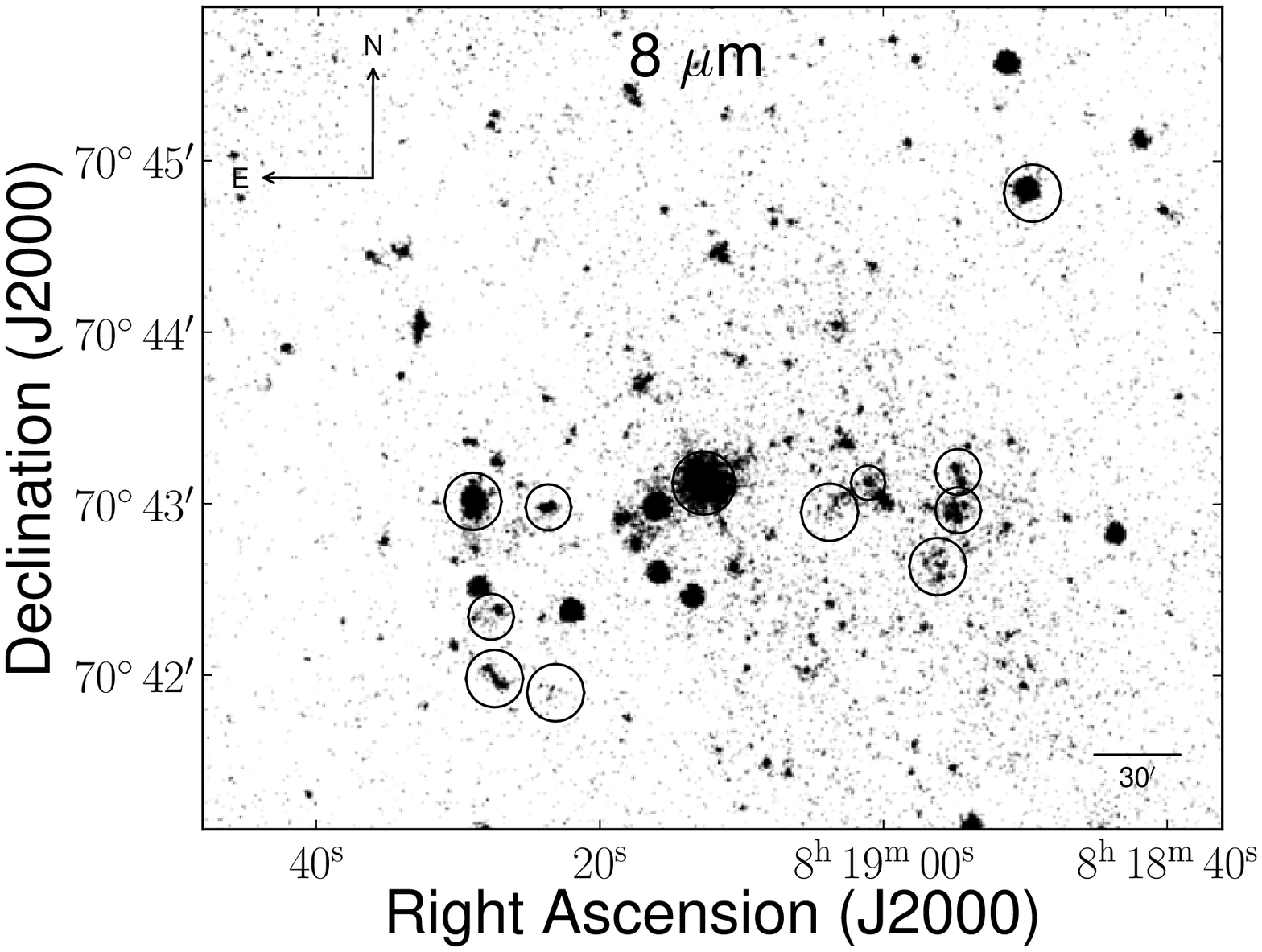}~\includegraphics[width=0.5\linewidth]{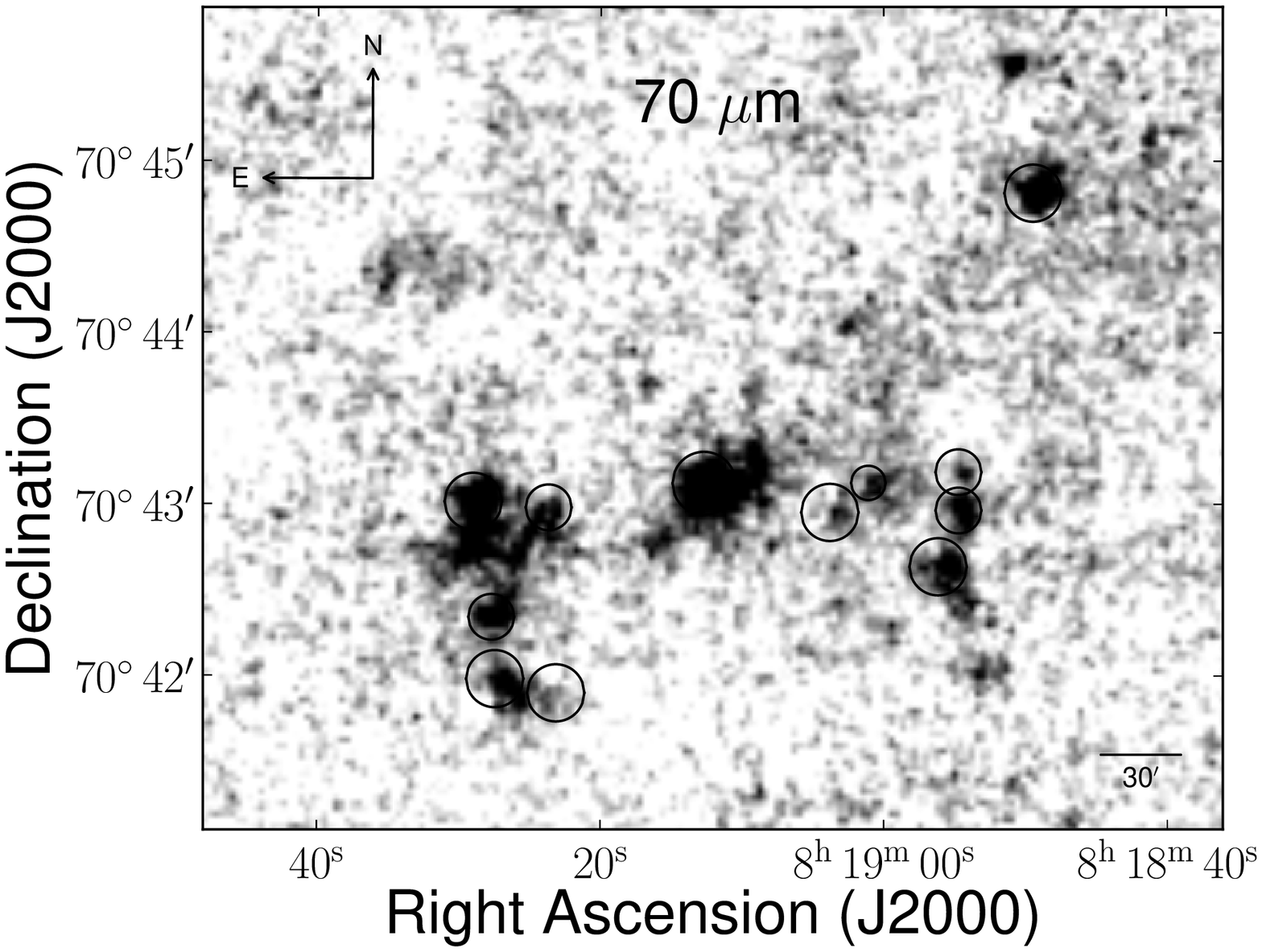}
\caption{Location of studied SFRs in the Ho~II galaxy identified according to Hodge et~al.~(1994): (a) the H$\alpha$ galaxy image from Kennicutt et~al.~(2003); (b) GALEX FUV image; (c), (d) 8-$\mu$m (left) and 70~$\mu$m (right) images from the Spitzer space
telescope. \hfill}
\end{figure*}

\section*{OBSERVATIONAL DATA}

A characteristic feature of the Ho~II galaxy that distinguishes it, for
example, from the Small Magellanic Cloud or the IC~10 galaxy is a nearly
complete absence of extended emission at 8~$\mu$m. The emission at 8~$\mu$m is
assumed to be generated mainly through the fluorescence of ionized PAH
molecules and, therefore, we may conclude that either the PAH molecules themselves or the UV~radiation needed to excite their IR bands is mostly absent
in Ho~II. The first alternative seems to be more likely, as diffuse
UV~radiation is visible in the GALEX images of the galaxy. In addition, the
galaxy is observed neither in CO emission (Leroy et~al.~2009) nor in
long-wavelength dust emission (Hunter et~al.~1986), suggesting the absence of
significant amounts of absorbing material that could shield the UV~radiation.

Nevertheless, there are isolated HII~complexes in Ho~II that are IR emission
sources. Khramtsova et~al.~(2013) considered the IR emission properties in
six HII~complexes in the galaxy Ho~II for which Moustakas et~al.~(2010)
provided metallicities. In more detail the galaxy's metallicity was studied by Egorov et~al.~(2013). As a result, in this study for our analysis we were able to select 12 complexes in active star-forming regions that are simultaneously visible in the 8~$\mu$m, 24~8~$\mu$m, and H$\alpha$ maps. Below, we use
designations of HII~complexes according to Hodge et~al. (1994); they were all
also included in the list by Stewart et~al.~(2000). The locations of the
chosen regions are shown in Fig.~1: (a) the H$\alpha$ map of the galaxy
obtained with the 2.1-m KPNO telescope (Kennicutt et~al.~2003), (b) the FUV map
obtained with the GALEX space telescope, and (c), (d) the IR~maps in the 8- and
70-$\mu$m bands.

To analyze SFRs in the Ho~II galaxy, we used the spectroscopic observations
performed with the 6-m SAO RAS telescope and presented in Egorov et~al.~(2013).
We retrieved the Spitzer observations at wavelengths from~3.6 to~24~$\mu$m from the SINGS\footnote{http://sings.stsci.edu} archive (Kennicutt et~al.~2003). Observational data at~70 and 160~$\mu$m are downloaded from the Herschel Science
Archive\footnote{http://herschel.esac.esa.int/Science\_Archive.shtml}, and the
UV~data are taken from the GALEX archive (in the FUV and NUV filters).

The procedures for data reduction and IR aperture photometry are described in
detail in Khramtsova et~al.~(2013), so here we present only
a brief description. The images from different instruments and at different
wavelengths are characterized by different point spread functions (PSFs). To
reduce all our data to a single angular resolution, we convolved them using the
programs described by Aniano et~al.~(2011). In addition, the data were converted to common physical units (Jy/pixel). The obtained images were used for aperture
photometry in all available filters. The aperture radii were chosen in
such a way that the SFR entirely fell within the aperture at all wavelengths.
We did not use apertures smaller than $12''$, which is the minimum
angular resolution. Since a typical SFR occupies no more than a few pixels
(especially at 160~$\mu$m), the contribution from incomplete pixels was taken
into account in the total flux. The background emission was subtracted for each
pixel within the aperture. A linear interpolation of the flux in the
surrounding pixels or, more precisely, in a ring with a width of several arcsec
was applied in the calculation.

To estimate the dust parameters, we used the model of Draine and Li~(2007). In
this model the dust IR~spectrum is computed under assumption that the radiation field in the investigated SFR is described by the expression

$$ u_\nu=Uu_\nu^{\textrm{MMP83}}, $$
where $U$ is the relative radiation intensity and $u_\nu^{\textrm{MMP83}}$ is
the radiation field (the mean intensity or energy density) in the solar
neighborhood determined by Mathis et~al.~(1983). It is also assumed in the
model that the mass of the dust $M_{\textrm{dust}}$ heated by a radiation field
in the range from $U_{\textrm{min}}$ to $U_{\textrm{max}}$ can be estimated
from the distribution
\begin{eqnarray*}
 {dM_{\textrm{dust}}\over
dU}=(1-\gamma)M_{\textrm{dust}}\delta(U-U_{\textrm{min}}) + \\
+ \gamma M_{\textrm{dust}}{\alpha-1\over
U_{\textrm{min}}^{1-\alpha}-U_{\textrm{max}}^{1-\alpha}}U^{-\alpha}.
\end{eqnarray*}
Here, $U_{\textrm{min}}$ is the minimum (background) radiation field in the
investigated object, $U_{\textrm{max}}$ is the maximum radiation field,
and $\delta$ is the delta~function. It was shown by Draine
et~al.~(2007) that results of calculations are only weakly dependent on $U_{\textrm{max}}$
and exponent~$\alpha$, so they can be fixed to reduce the number of free
parameters. Based on observational data for the Milky Way, Draine et~al.~(2007) proposed
values of $10^6$ and~2, respectively, for these parameters. The mass fraction of dust illuminated
by an enhanced radiation field (greater than $U_{\textrm{min}}$) is $\gamma$.
The dust is assumed to be described by the size distribution from Weingartner and
Draine~(2001) with a PAH mass fraction $q_{\textrm{PAH}}$.

Using the least-squares method, for each SFR we found the model that best fits the
observations at 3.6, 4.5, 5.8, 8.0, 24, 70, 100, and 160~$\mu$m,
determining the following parameters of the SFR: the
PAH mass fraction $q_{\textrm{PAH}}$, the minimum intensity of the dust-heating
radiation field, the fraction of the dust heated by an enhanced radiation
field, and the dust mass. The observations at 3.6~$\mu$m were used to subtract
the stellar background by the technique described in Marble et~al.~(2010). The
Monte Carlo method was used to determine the parameter estimation errors.

Using on our aperture photometry, we calculated the following flux ratios:
$$ P_{8}=\frac{\nu F^{\textrm{ns}}_{\nu}(8\textrm{~$\mu$m})}{\nu
F_{\nu}(70\textrm{~$\mu$m})+\nu F_{\nu}(160\textrm{~$\mu$m})},
$$
$$
P_{24}=\frac{\nu F^{\textrm{ns}}_{\nu}(24\textrm{~$\mu$m})}{\nu
F_{\nu}(70\textrm{~$\mu$m})+\nu F_{\nu}(160\textrm{~$\mu$m})},
$$
$$
R_{71}=\frac{\nu F_{\nu}(70\textrm{~$\mu$m})}{\nu
F_{\nu}(160\textrm{~$\mu$m})}.
$$
Here, $\nu$ denotes the central frequency of the filter. The superscript
``ns'' denotes the flux from which the stellar contribution was subtracted. This
component was subtracted from the fluxes at~8 and 24~$\mu$m. The results of the
UV photometry are presented as FUV and NUV magnitudes. Apart from
the metallicities,emission line intensities were also taken from Egorov
et~al.~(2013). The results of the photometry are presented in Table~1.

\begin{table*}[t!]
%%% Table:1
\caption{Results of the photometry for SFRs in the Holmberg~II galaxy (the numbers in parentheses indicate powers of 10)}
\tiny{\begin{tabular}{l|c|c|c|c|c|c|c}
\hline
Region  &   $F_8$              &     $F_{24}$           & $F_{70}$    & $F_{100}$         & $F_{160}$         &    FUV         &  NUV \\
\hline
HSK7     & $6.6({-}4)\pm2.2({-}3)$ & $8.7({-}3)\pm1.9({-}4)$ & $6.9({-}2)\pm2.7({-}3)$ & $6.8({-}2)\pm3.9({-}3)$ & $3.7({-}2)\pm1.9({-}3)$ & $16.85\pm0.08$ & $16.81\pm0.04$\\
HSK15    & $6.9({-}5)\pm1.3({-}5)$ & $6.7({-}4)\pm8.5({-}5)$ & $5.9({-}3)\pm2.2({-}3)$ & $8.6({-}3)\pm2.6({-}3)$ & $3.7({-}3)\pm2.4({-}3)$ & $19.93\pm0.33$ & $20.14\pm0.36$\\
HSK16--17& $1.2({-}4)\pm8.9({-}6)$ & $1.1({-}3)\pm6.0({-}5)$ & $1.7({-}2)\pm1.9({-}3)$ & $2.2({-}2)\pm1.8({-}3)$ & $1.3({-}2)\pm1.5({-}3)$ & $17.37\pm0.11$ & $17.45\pm0.11$\\
HSK20    & $9.4({-}5)\pm1.0({-}5)$ & $8.1({-}4)\pm6.1({-}5)$ & $2.6({-}2)\pm3.2({-}3)$ & $3.1({-}2)\pm3.7({-}3)$ & $1.8({-}2)\pm3.0({-}3)$ & $17.48\pm0.27$ & $17.42\pm0.25$\\
HSK25    & $2.2({-}4)\pm1.4({-}5)$ & $7.1({-}4)\pm8.2({-}5)$ & $1.3({-}2)\pm1.4({-}3)$ & $2.2({-}2)\pm3.6({-}3)$ & $2.2({-}2)\pm3.8({-}3)$ & $19.38\pm0.42$ & $19.24\pm0.37$\\
HSK31    & $8.6({-}5)\pm2.1({-}5)$ & $4.1({-}4)\pm6.5({-}5)$ & $7.0({-}3)\pm2.2({-}3)$ & $1.2({-}2)\pm2.6({-}3)$ & $1.3({-}2)\pm2.7({-}3)$ & $19.34\pm0.20$ & $19.64\pm0.26$\\
HSK45    & $6.1({-}3)\pm3.1({-}4)$ & $2.1({-}2)\pm3.2({-}4)$ & $1.1({-}1)\pm6.4({-}3)$ & $1.2({-}1)\pm7.1({-}3)$ & $6.9({-}2)\pm4.5({-}3)$ & $15.06\pm0.02$ & $15.33\pm0.05$\\
HSK59    & $8.0({-}5)\pm1.1({-}5)$ & $1.2({-}3)\pm1.3({-}4)$ & $7.1({-}3)\pm3.2({-}3)$ & $6.3({-}3)\pm3.0({-}3)$ & $5.3({-}3)\pm2.0({-}3)$ & $18.13\pm0.06$ & $18.19\pm0.10$\\
HSK61    & $2.8({-}5)\pm9.3({-}6)$ & $2.0({-}4)\pm8.5({-}5)$ & $1.2({-}2)\pm3.6({-}3)$ & $2.1({-}2)\pm4.0({-}3)$ & $1.4({-}2)\pm2.7({-}3)$ & $18.07\pm0.15$ & $18.06\pm0.14$\\
HSK67    & $1.3({-}3)\pm1.1({-}4)$ & $2.1({-}3)\pm1.8({-}4)$ & $2.7({-}2)\pm2.4({-}3)$ & $2.3({-}2)\pm2.9({-}3)$ & $1.5({-}2)\pm1.2({-}3)$ & $17.28\pm0.14$ & $17.43\pm0.14$\\
HSK70    & $1.2({-}3)\pm3.9({-}5)$ & $1.1({-}3)\pm7.9({-}5)$ & $1.5({-}2)\pm1.8({-}3)$ & $1.7({-}2)\pm2.8({-}3)$ & $7.0({-}3)\pm1.7({-}3)$ & $15.74\pm0.04$ & $15.89\pm0.06$\\
HSK71--73& $1.2({-}2)\pm2.7({-}4)$ & $2.2({-}2)\pm4.4({-}4)$ & $7.7({-}2)\pm6.5({-}3)$ & $6.6({-}2)\pm6.6({-}3)$ & $3.1({-}2)\pm3.7({-}3)$ & $16.50\pm0.14$ & $16.54\pm0.13$\\
\hline
\end{tabular}}
\end{table*}

\section*{THE AGES OF STAR-FORMING REGIONS}

As was shown by Copetti et~al.~(1986), the equivalent width of the
$\textrm{H}{\beta}$ emission line correlates well with an age of an HII~region,
while being virtually independent of the electron temperature of the medium.
This correlation arises because the number of ionizing photons from
massive stars decreases with age, while the continuum level from low-mass stars at H${\beta}$
wavelength increases. The possibility of using
EW(H${\beta}$) as an age indicator was demonstrated in several studies (see,
e.g., Schaerer and Vacca~1998; Leitherer et~al.~1999).

In this paper, to estimate ages we use the grid of models presented by Levesque et~al.~(2010)
based on the evolutionary tracks from Schaerer and Vacca~(1998) for various
metallicities. However, this method has several
shortcomings described in detail by Stasi\'nska and Leitherer~(1996), with the
main one being the assumption about a single stellar population in a given HII~complex.
If there are stars of previous generations in the HII~complex, then they can
raise the continuum level, thereby reducing the H${\beta}$ equivalent width
and, accordingly, overestimating the age. In addition, internal extinction in
the HII~complex can decrease the number of ionizing photons. However,
the dust mass in Ho~II is insignificant and extinction should not affect
our results significantly.

To assess the correctness of the SFR age estimates, we used observations
of Ho~II performed at the 6-m SAO RAS telescope with a Fabry--Perot
interferometer (FPI) in $\textrm{H}{\alpha}$ and [SII]~6717~\AA\ lines. (A
detailed study of the ionized and neutral gas kinematics in the galaxy will be
presented in the paper by Egorov, Lozinskaya, and Moiseev being prepared for
publication.) The derived data cubes were used to estimate the expansion
velocities of SFRs and to determine their kinematic ages. We attempted to
determine the expansion velocity from the characteristic splitting of the
position--velocity (PV) diagrams for shell-like HII complexes, but no such evidence
was revealed in most of the discussed regions. Their expansion is only revealed by the increasing line FWHM toward the center of the
region. To estimate the expansion velocity roughly, in each complex we identified locations 
of the radial velocity minimum and maximum and assumed that the
minimum and maximum velocities correspond to the emission from bright
parts of the approaching and receding sides of the shell, respectively. The
results are presented in Table~2. For regions where the inferred
expansion velocity turned out to be smaller than the FPI resolution
(18~km~s$^{-1}$), in Table~2 we give an upper limit for the expansion velocity equal to
9~km~s$^{-1}$, while the value formally found from our data is given in
parentheses. The value of 9~km~s$^{-1}$ was also assumed in cases where we
failed to obtain an estimate of the expansion velocity. Typical uncertainty of expansion velocity measurements is 2~km~s$^{-1}$.

\begin{table}[t!]
%%% Table:2
\caption{Comparison between ages determined from the H$\beta$ equivalent width
and kinematic ages} 

\small{\begin{tabular}{l|c|c|c|c}
\hline \multicolumn{1}{c|}{Region} &
\parbox[c][1.0cm]{0.6cm}{$V_{\textrm{exp}}$, km~s$^{-1}$} &
\multicolumn{1}{c|}{\parbox[c][1.0cm]{1.2cm}{Diameter, arcsec}} &
\parbox[c][1.0cm]{0.5cm}{Age, Myr}  &
\parbox[c][1.0cm]{0.6cm}{Age ($\textrm{H}\beta$),\\ Myr} \\
\hline
HSK15 & $<9$ & 2.8 & $>1.5$ & 4.8 \\
HSK16--17 & $<9 (6)$ & 6.9 & $>2.3 (5.6)$ & 5.3 \\
HSK20 & $<9$ & 4.0 & $>2.2$ & 5.3 \\
HSK25 & $<9$ & 4.6 & $>2.5$ & 6.2 \\
HSK31 & $<9 (4)$ & 6.2 & $>3.4 (7.6)$ & 6.3 \\
HSK45 & 16 & 11.3 & $>3.5$ & 3.7 \\
HSK59 & $<9 (4)$ & 2.0 & $>1.1 (2.5)$ & 4.1 \\
HSK61 & $<9 (5)$ & 3.8 & $>2.1 (3.7)$ & 7.7 \\
HSK67 & $<9 (6)$ & 4.3 & $>2.4 (3.5)$ & 3.8 \\
HSK70 & $<9 (6)$ & 6.3 & $>3.4 (5.2)$ & 4.8 \\
HSK71 & $<9$ & 4.7 & $>2.6$ & 3.5 \\
HSK73 & 10 & 7.5 & $>3.7$ & 3.5 \\
\hline
\end{tabular}}
\end{table}

\begin{figure}[b!]
%%% Figure:2
\includegraphics[width=\linewidth]{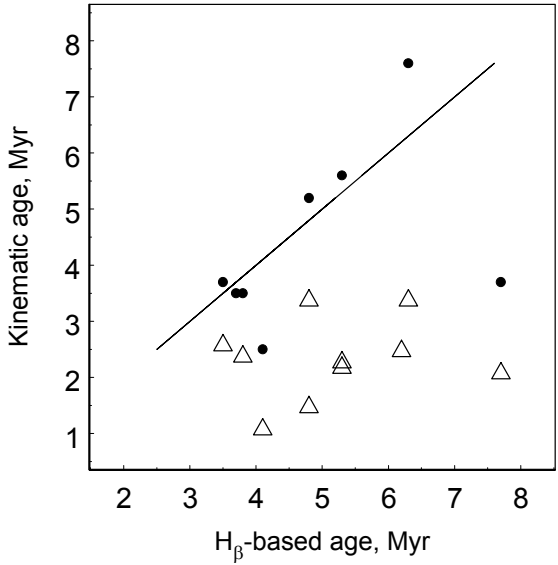}

\caption{Comparison of ages determined from the H$\beta$
equivalent width and the kinematic ages of SFRs. The black circles
and triangles indicate formal kinematic age estimates and lower age limits, respectively. The straight line corresponds
to the equality of the two age estimates. \hfill}
\end{figure}

\begin{table*}[t!]
%%% Table:3
\caption{Model parameters of SFRs in the Holmberg II galaxy}
\tiny{\begin{tabular}{l|c|c|c|c|c|c|c|c|c}
\hline \multicolumn{1}{c|}{Region} & 12~$+$~log(O/H)&
$q_{\textrm{PAH}}$, ${\%}$ & $U_{\min}$ & $\gamma$     &
$M_{\textrm{dust}}$, $M_\odot$& \parbox[c][1cm]{0.5cm}{Age,
Myr}&
\parbox[c][1.2cm]{1cm}{Age group (Stewart et~al.~2000)}& $F_8/F_{24}$ & [OIII]/H$\beta$ \\
\hline
HSK7     & $7.76\pm0.03$ & $0.41\pm0.04$ & $23.51\pm0.87$ & $6.6({-}2)\pm1.1({-}2)$ & $31.8\pm15.2$ & $3.7\pm0.1$ &\phantom{0.}0--3.5   & $0.050\pm0.002$ & $4.3\pm0.2$\phantom{0}\\
HSK15    & $7.63\pm0.10$ & $0.48\pm0.01$ & $17.99\pm3.35$ & $4.2({-}2)\pm1.2({-}2)$ & $ 4.1\pm 0.7$ & $4.8\pm0.1$ & 3.5--4.5 & $0.021\pm0.023$ & $1.6\pm0.1$\phantom{0}\\
HSK16--17& $7.68\pm0.05$ & $0.48\pm0.04$ & $13.56\pm0.55$ & $2.0({-}2)\pm8.0({-}3)$ & $16.0\pm 2.2$\phantom{0} & $5.3\pm0.2$ & 3.5--4.5 & $0.062\pm0.010$ & $2.1\pm0.1$\phantom{0}\\
HSK20    & $7.45\pm0.04$ & $0.45\pm0.07$ & $20.57\pm0.29$ & $3.0({-}3)\pm2.0({-}3)$ & $16.8\pm 8.6$\phantom{0} & $5.3\pm0.2$ & 4.5--6.3 & $0.079\pm0.016$ & $1.0\pm0.1$\phantom{0}\\
HSK25    & $7.53\pm0.02$ & $1.12\pm0.52$ & \phantom{0}$ 2.76\pm0.14$ & $1.5({-}2)\pm7.0({-}3)$ & $73.8\pm36.6$ & $6.2\pm0.2$ & 4.5--6.3 & $0.181\pm0.032$ & $1.1\pm0.02$\\
HSK31    & $7.97\pm0.12$ & $0.91\pm0.04$ & \phantom{0}$ 2.09\pm0.35$ & $1.6({-}2)\pm6.0({-}3)$ & $53.4\pm17.6$ & $6.3\pm0.2$ & 4.5--6.3 & $0.149\pm0.058$ & $3.3\pm0.04$\\
HSK45    & $7.89\pm0.03$ & $2.95\pm0.22$ & $16.49\pm1.90$ & $1.1({-}1)\pm4.4({-}2)$ & $75.9\pm16.6$ & $3.7\pm0.1$ & \phantom{0.}0--3.5  & $0.199\pm0.009$ & $3.6\pm0.02$\\
HSK59    & $7.60\pm0.03$ & $0.56\pm0.13$ & \phantom{0}$ 6.84\pm1.41$ & $1.1({-}1)\pm5.1({-}2)$ & $ 8.0\pm 4.7$ & $4.1\pm0.1$ & 3.5--4.5 & $0.059\pm0.012$ & $1.8\pm0.02$\\
HSK61    & $7.66\pm0.03$ & $0.50\pm0.04$ & \phantom{0}$ 3.06\pm0.64$ & $1.0({-}4)\pm9.9({-}5)$ & $51.6\pm 6.2$\phantom{0} & $7.7\pm0.2$ & 3.5--4.5 & $0.084\pm0.070$ & $2.1\pm0.03$\\
HSK67    & $7.57\pm0.01$ & $0.40\pm0.12$ & $10.46\pm1.46$ & $4.5({-}2)\pm8.0({-}3)$ & $24.4\pm 7.0$\phantom{0} & $3.8\pm0.1$ & \phantom{0.}0--3.5  & $0.035\pm0.006$ & $3.0\pm0.02$\\
HSK70    & $7.72\pm0.01$ & $0.44\pm0.10$ & $25.00\pm1.28$ & $2.9({-}2)\pm7.0({-}3)$ & $ 7.3\pm 2.9$ & $4.8\pm0.1$ & 3.5--4.5 & $0.037\pm0.066$ & $2.9\pm0.02$\\
HSK71--73& $7.72\pm0.04$ & $0.48\pm0.12$ & $25.00\pm1.59$ & $2.3({-}1)\pm3.8({-}2)$ & $23.3\pm 5.4$\phantom{0} & $3.5\pm0.1$ & \phantom{0.}0--3.5  & $0.038\pm0.001$ & $2.2\pm0.02$\\
\hline
\end{tabular}}
\end{table*}

The kinematic age was estimated from the model of Mac~Low and McCray~(1988).
For an expanding supershell driven by the combined action of supernovae and
stellar winds, we used the relation
$$
t\approx0.6 R/V_{\textrm{exp}},
$$
where $R$ is the radius in parsecs, $V_{\textrm{exp}}$ is the expansion
velocity in~km~s$^{-1}$, and~$t$ is the age in~Myr. As can be seen from
Table~2, we were able to estimate reliably the kinematic age only in two considered regions, HSK45 and HSK73. For a few more regions, we obtained rough
estimates.

Overall, given the mentioned errors in the kinematic age estimates and the
model dependence of the estimates based on the H$\beta$ equivalent width, Fig.~2
shows an agreement between the results obtained with the two methods. The only exception is
the HSK61 region in which the H$\beta$ line has the smallest equivalent width
corresponding to an age of more than 7~Myr constitutes.

\begin{figure*}[t!]
%%% Figure:3
\center{~~~~(a)~~~~~~~~~~~~~~~~~~~~~~~~~~~~~~~~~~~~~~~~~~(b)~~~~~~~~~~~~~~~~~
~~~~~~~~~~~~~~~~~~~~~~~~~(c)}

\includegraphics[width=0.33\linewidth]{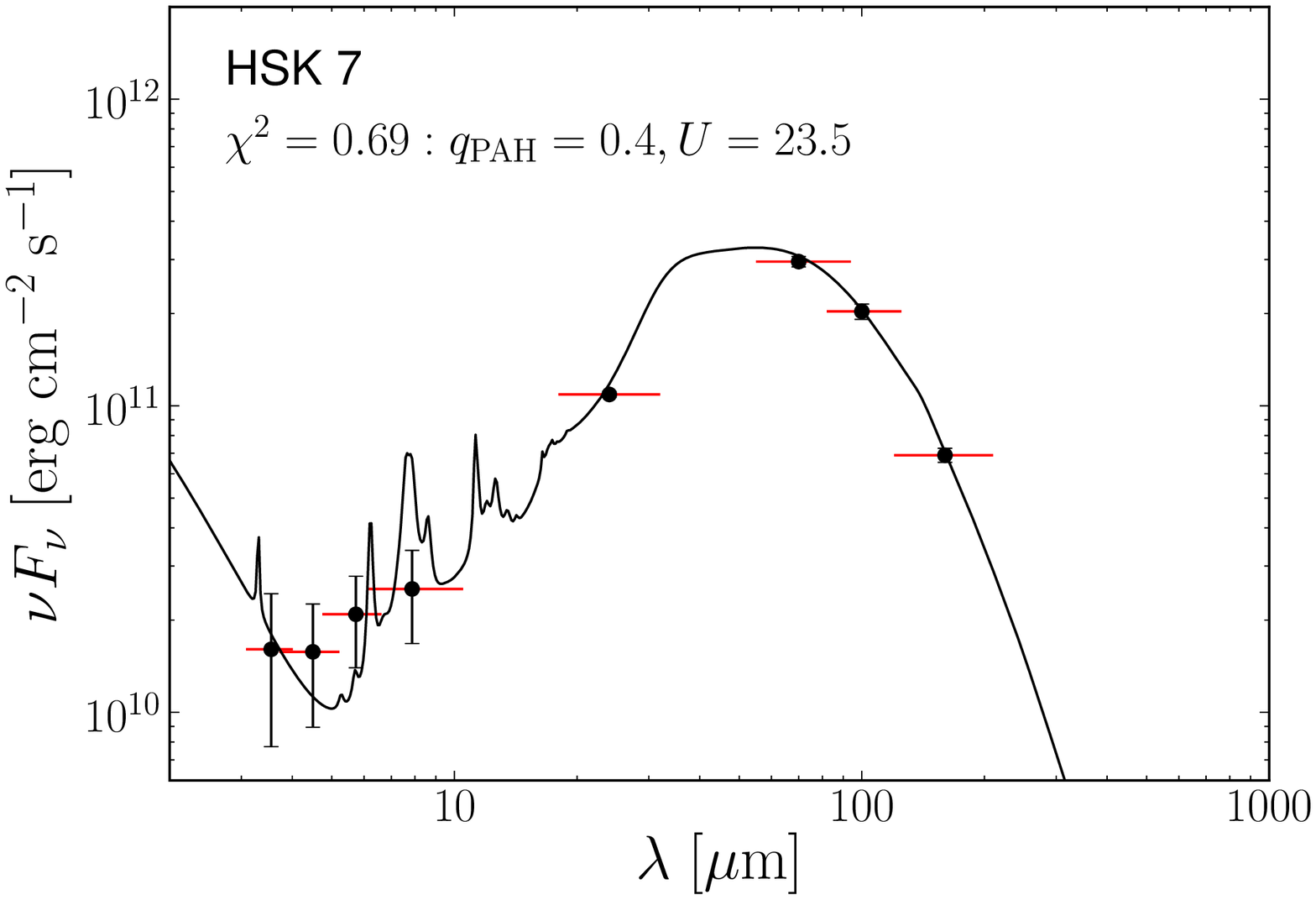}~\includegraphics[width=0.33\linewidth]{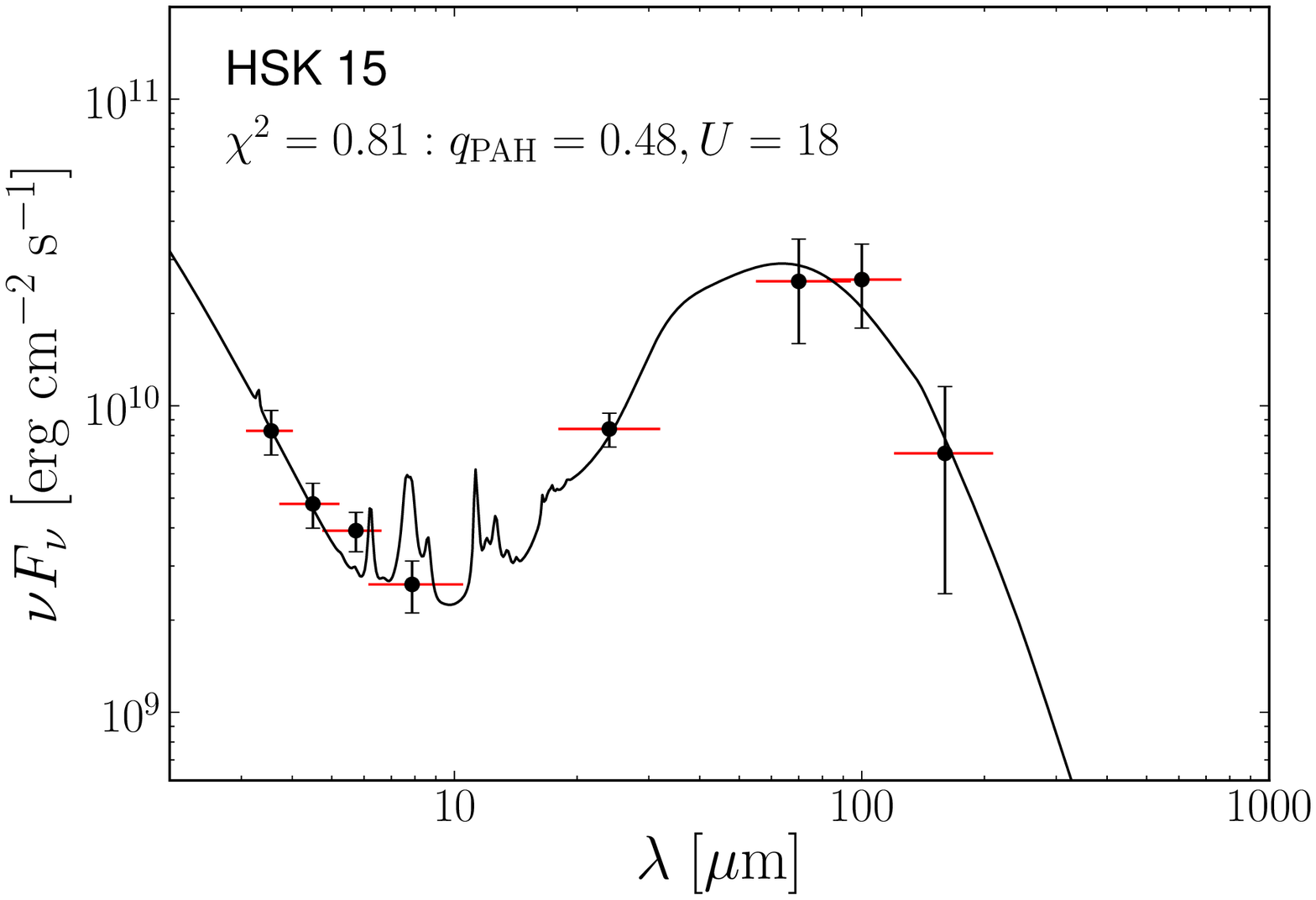}~\includegraphics[width=0.33\linewidth]{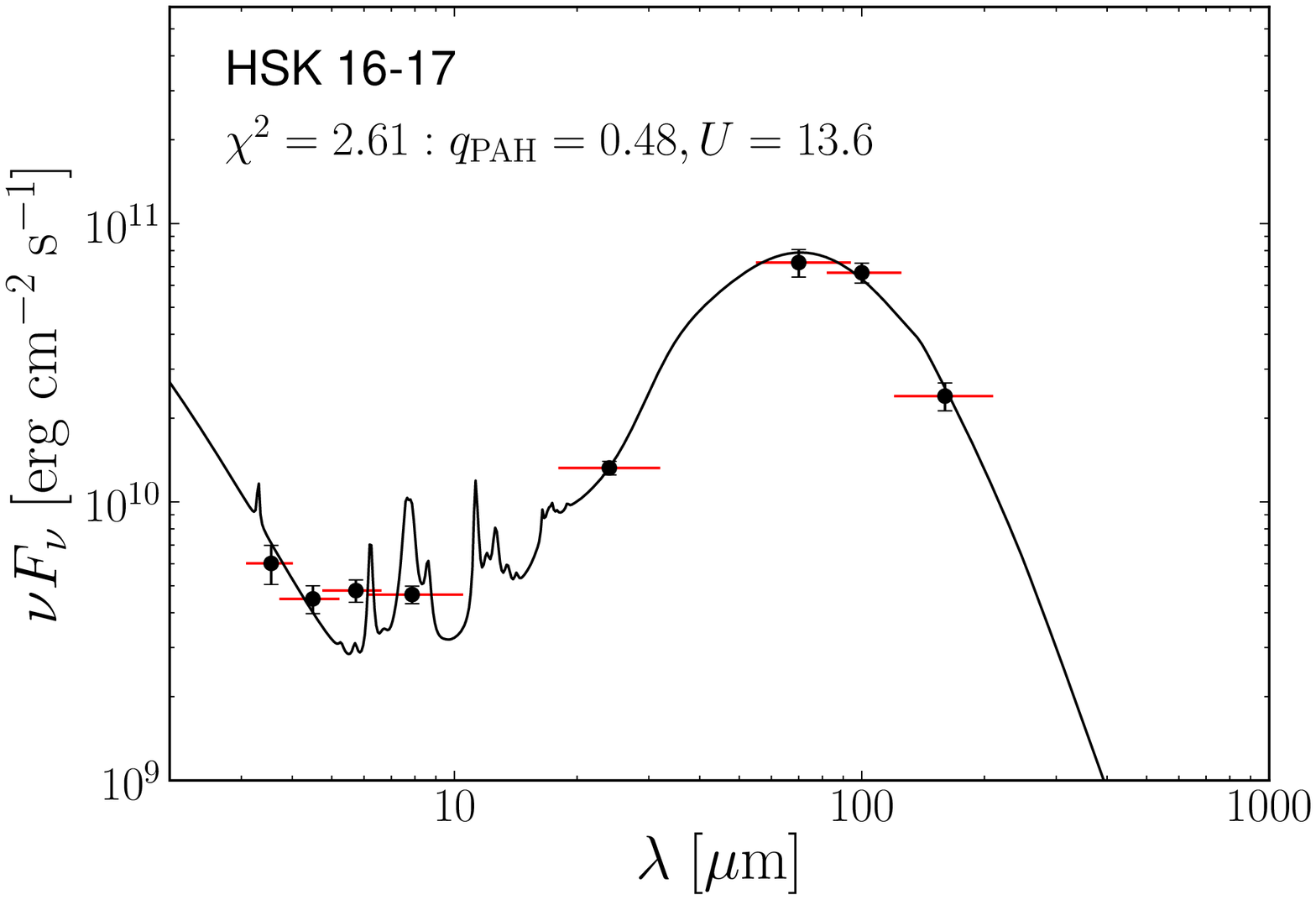}

\center{~~~~(d)~~~~~~~~~~~~~~~~~~~~~~~~~~~~~~~~~~~~~~~~~~(e)~~~~~~~~~~~~~~~~~
~~~~~~~~~~~~~~~~~~~~~~~~~(f)}

\includegraphics[width=0.33\linewidth]{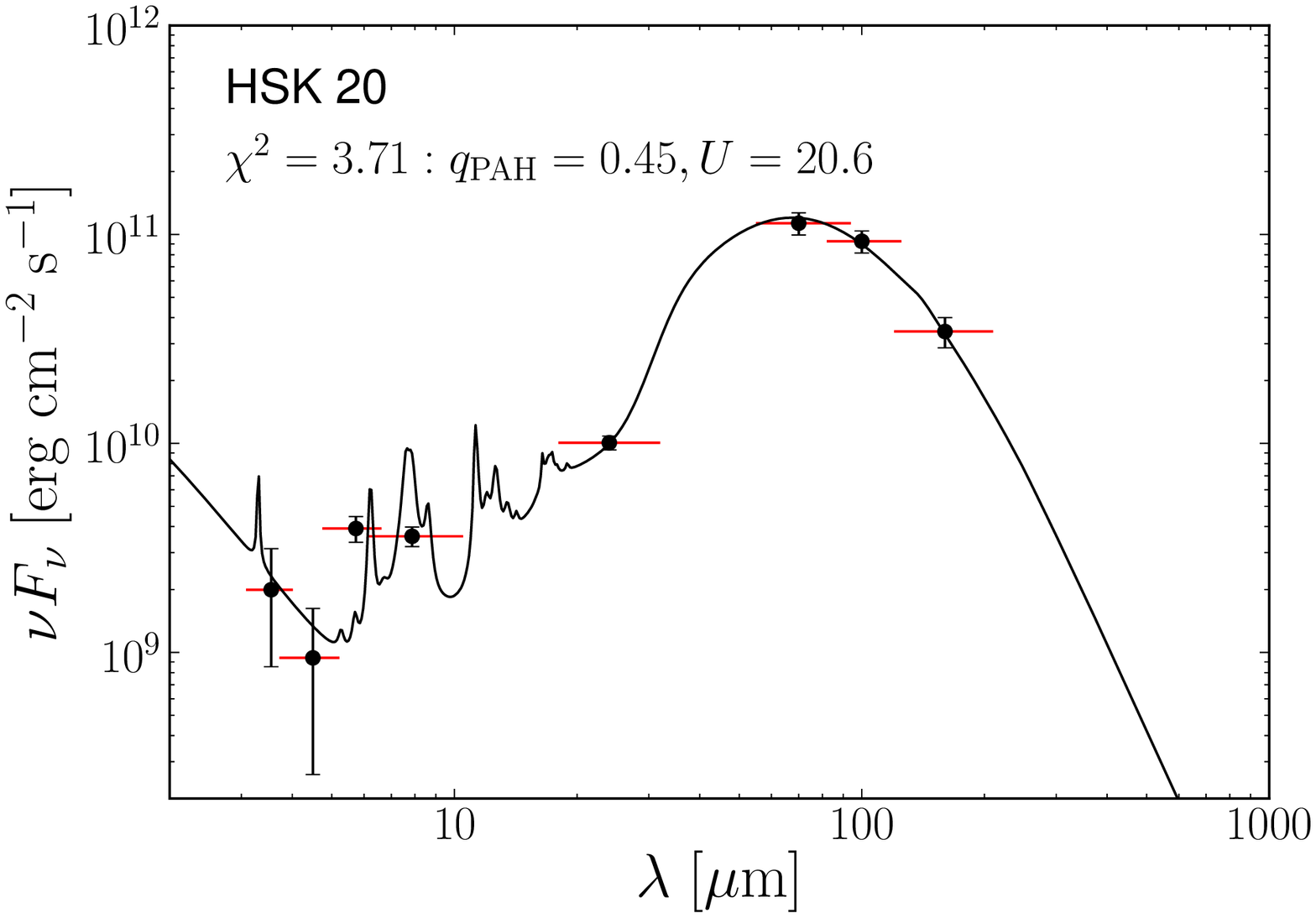}~\includegraphics[width=0.33\linewidth]{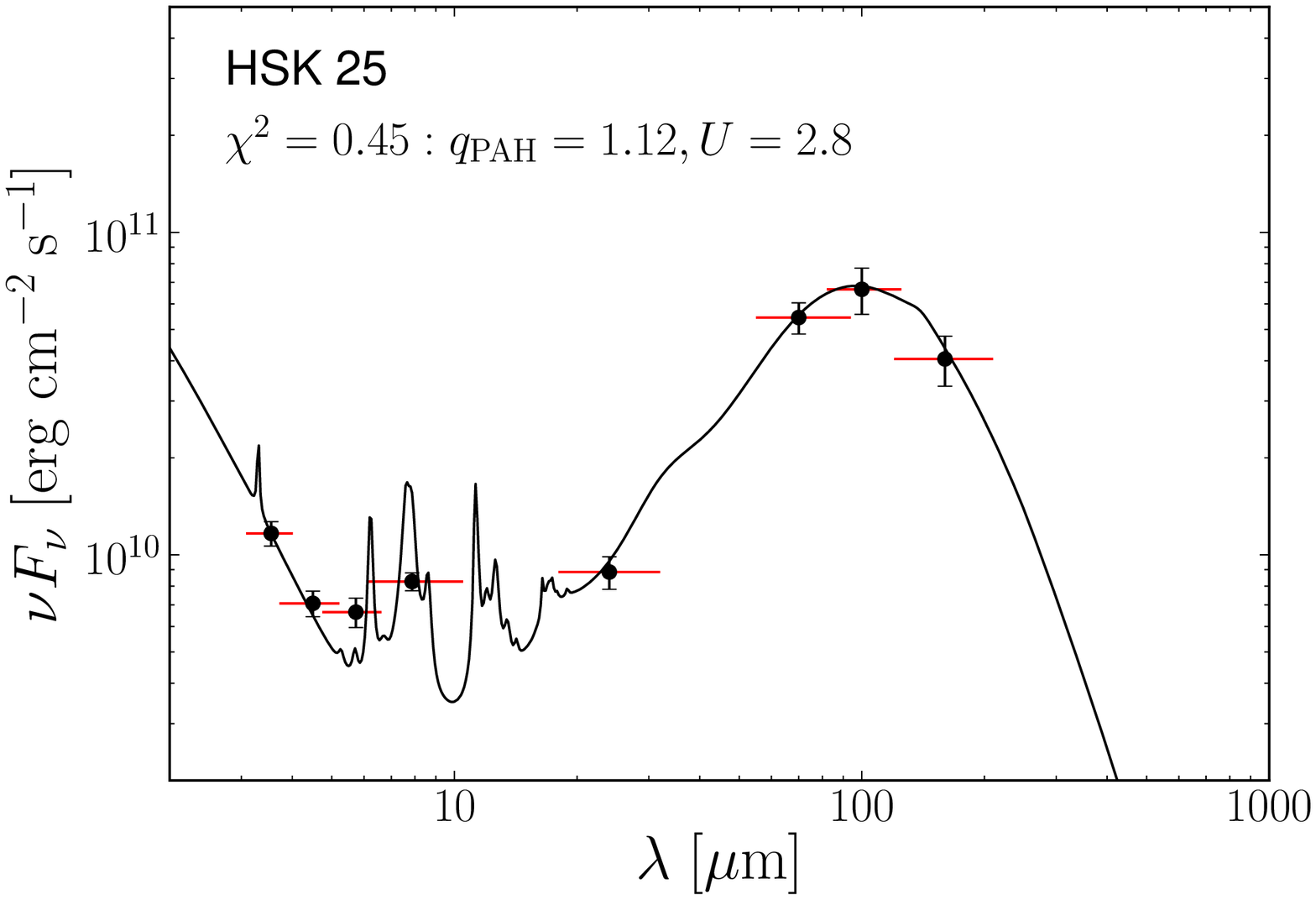}~\includegraphics[width=0.33\linewidth]{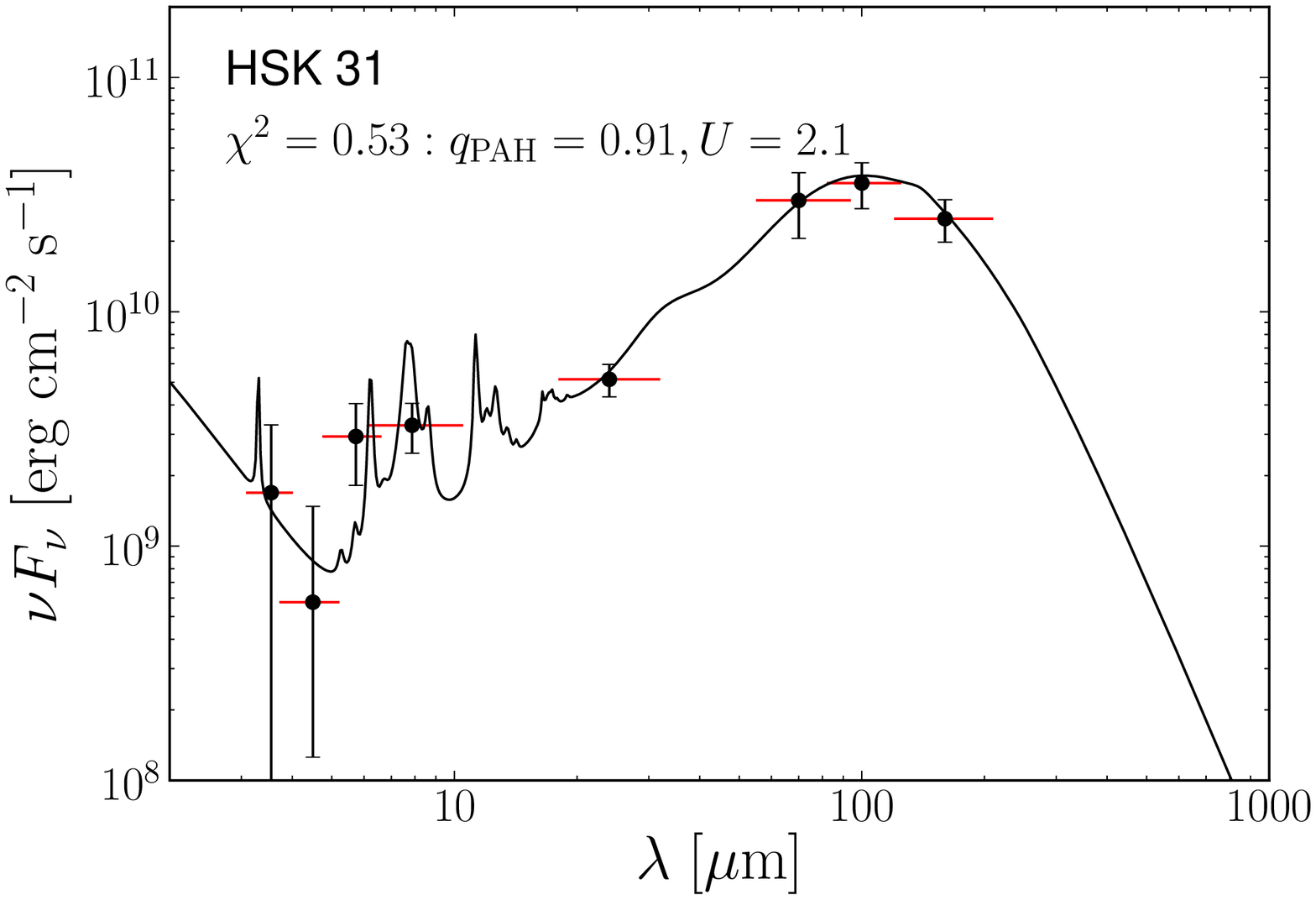}

\center{~~~~(g)~~~~~~~~~~~~~~~~~~~~~~~~~~~~~~~~~~~~~~~~~~(h)~~~~~~~~~~~~~~~~~
~~~~~~~~~~~~~~~~~~~~~~~~~(j)}

\includegraphics[width=0.33\linewidth]{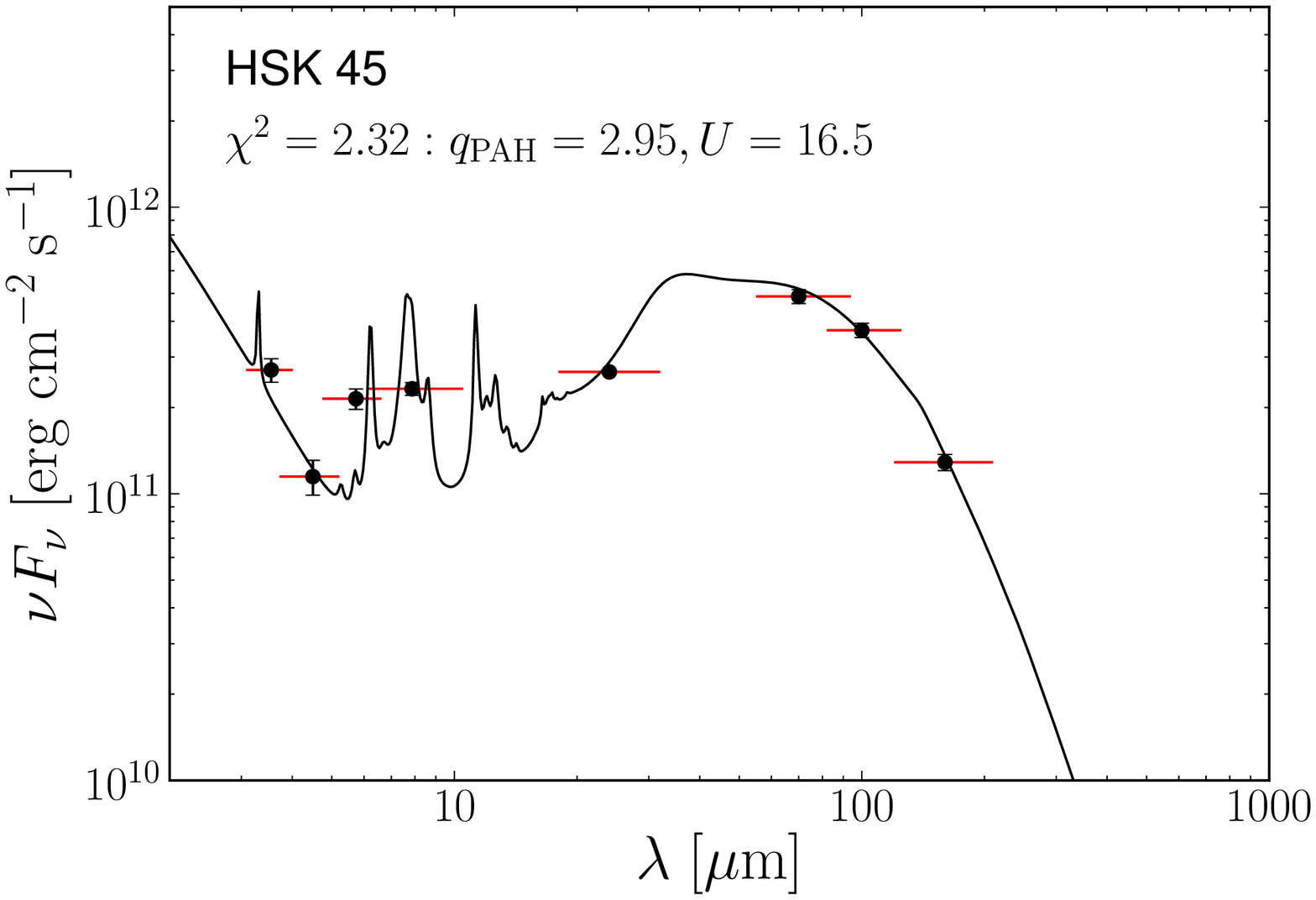}~\includegraphics[width=0.33\linewidth]{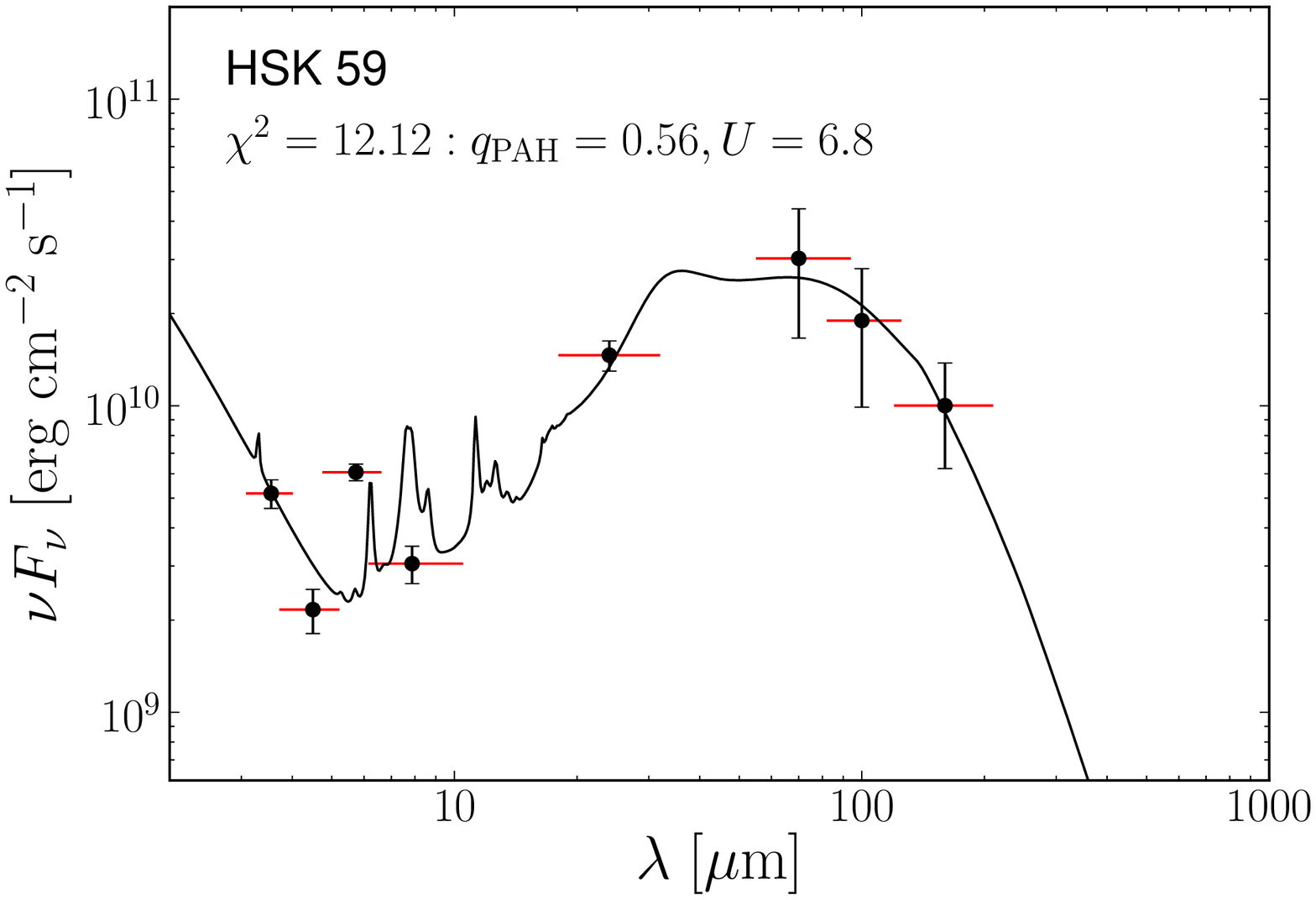}~\includegraphics[width=0.33\linewidth]{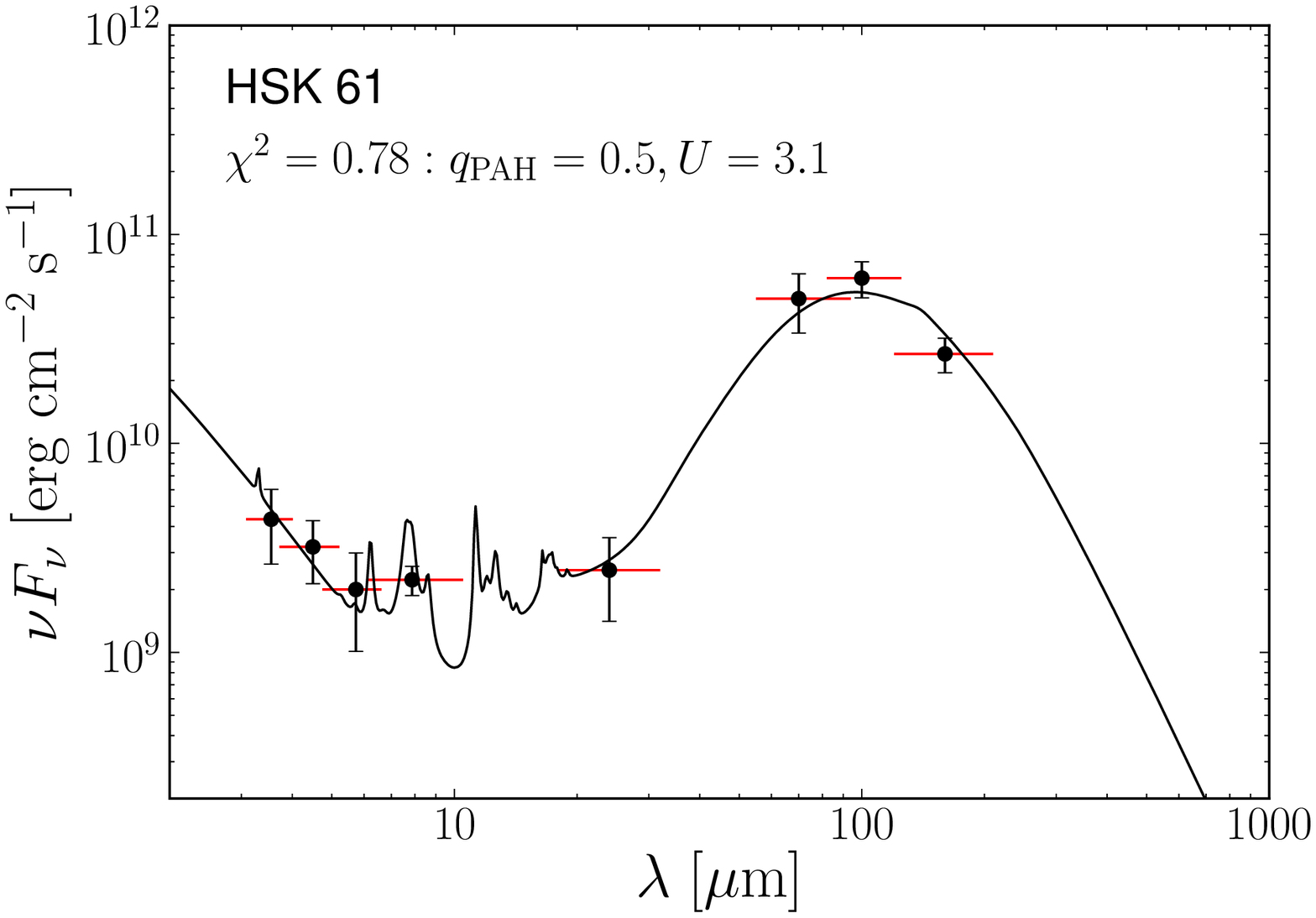}

\center{~~~~(i)~~~~~~~~~~~~~~~~~~~~~~~~~~~~~~~~~~~~~~~~~~(k)~~~~~~~~~~~~~~~~~
~~~~~~~~~~~~~~~~~~~~~~~~~(l)}

\includegraphics[width=0.33\linewidth]{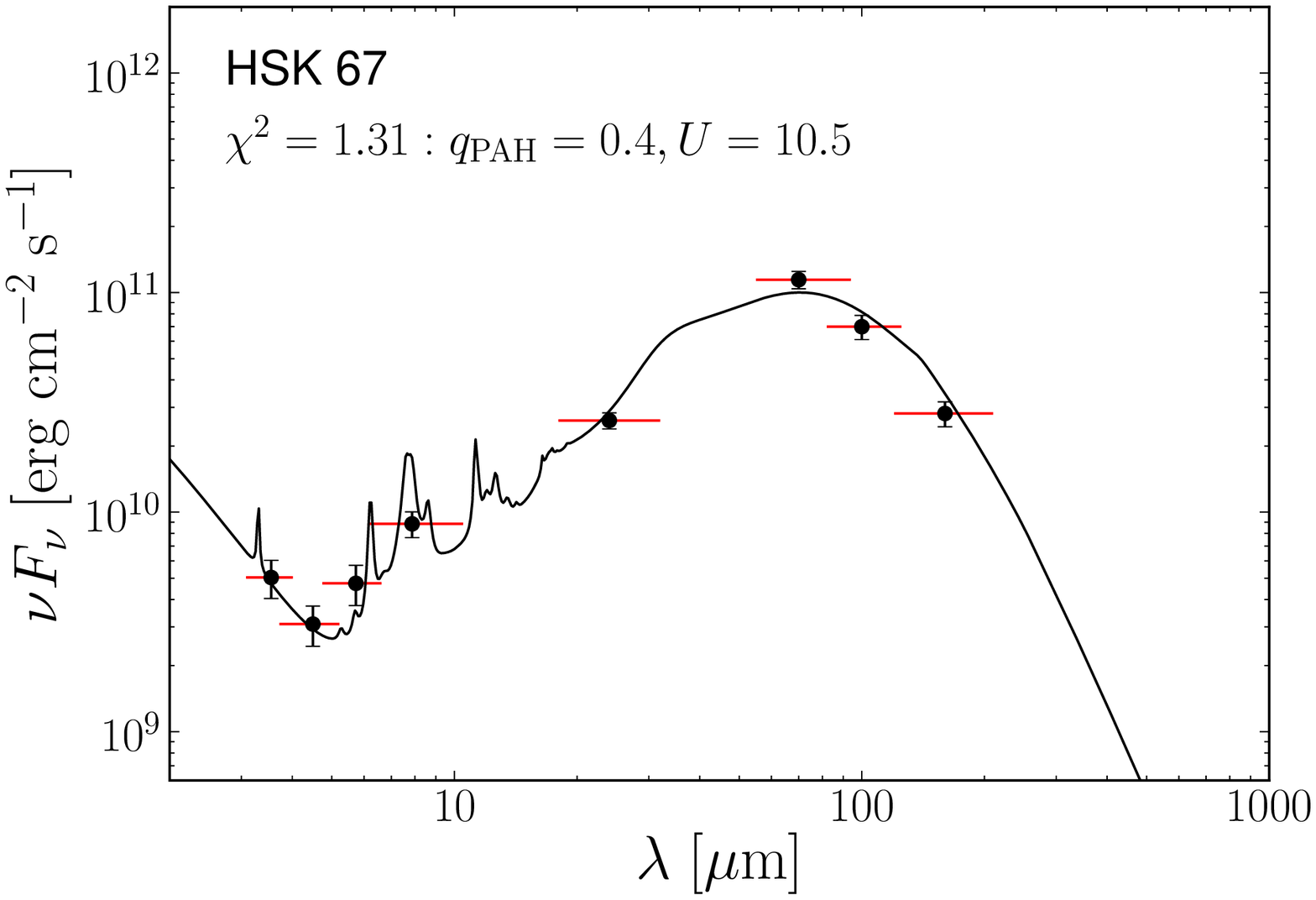}~\includegraphics[width=0.33\linewidth]{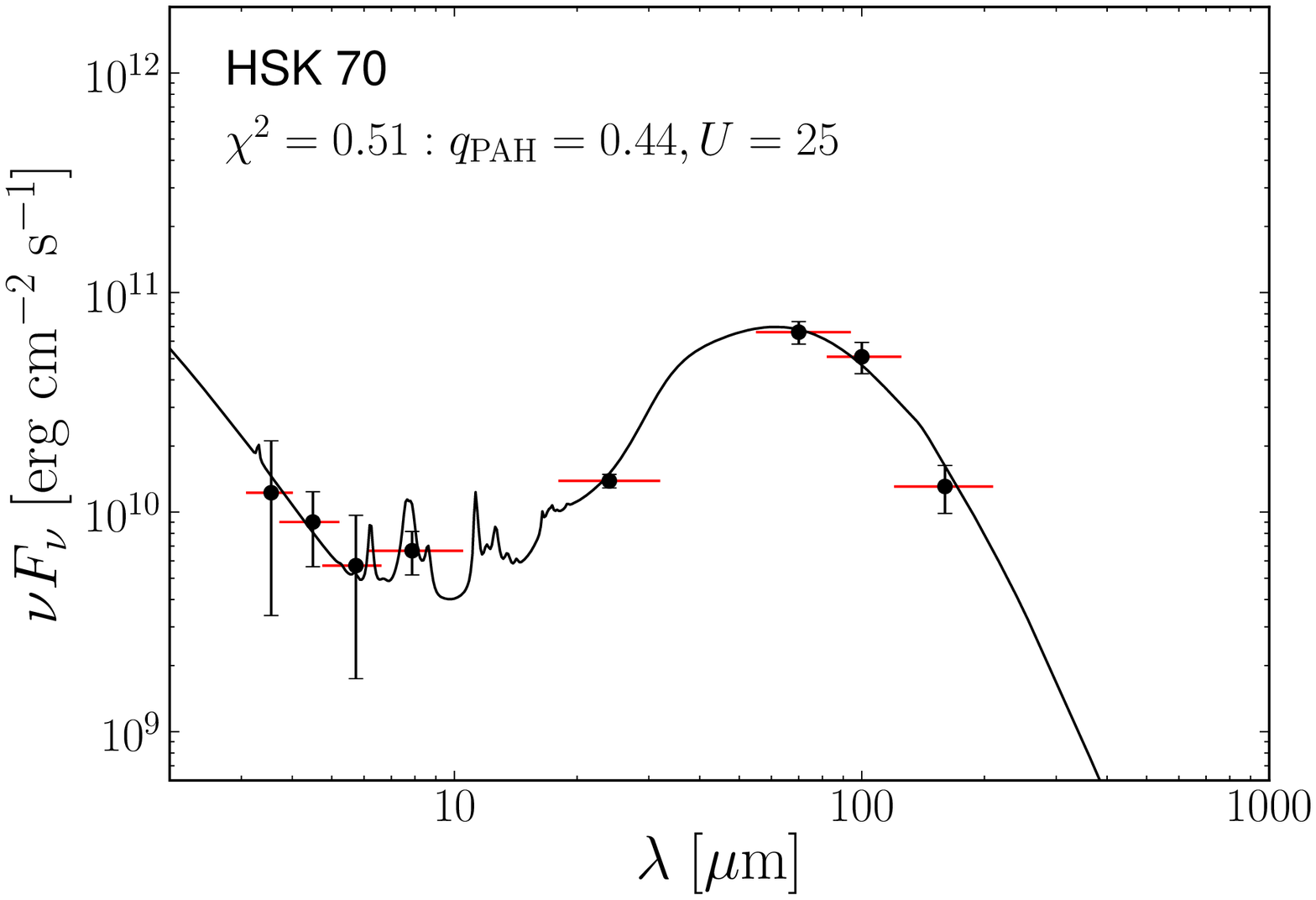}~\includegraphics[width=0.33\linewidth]{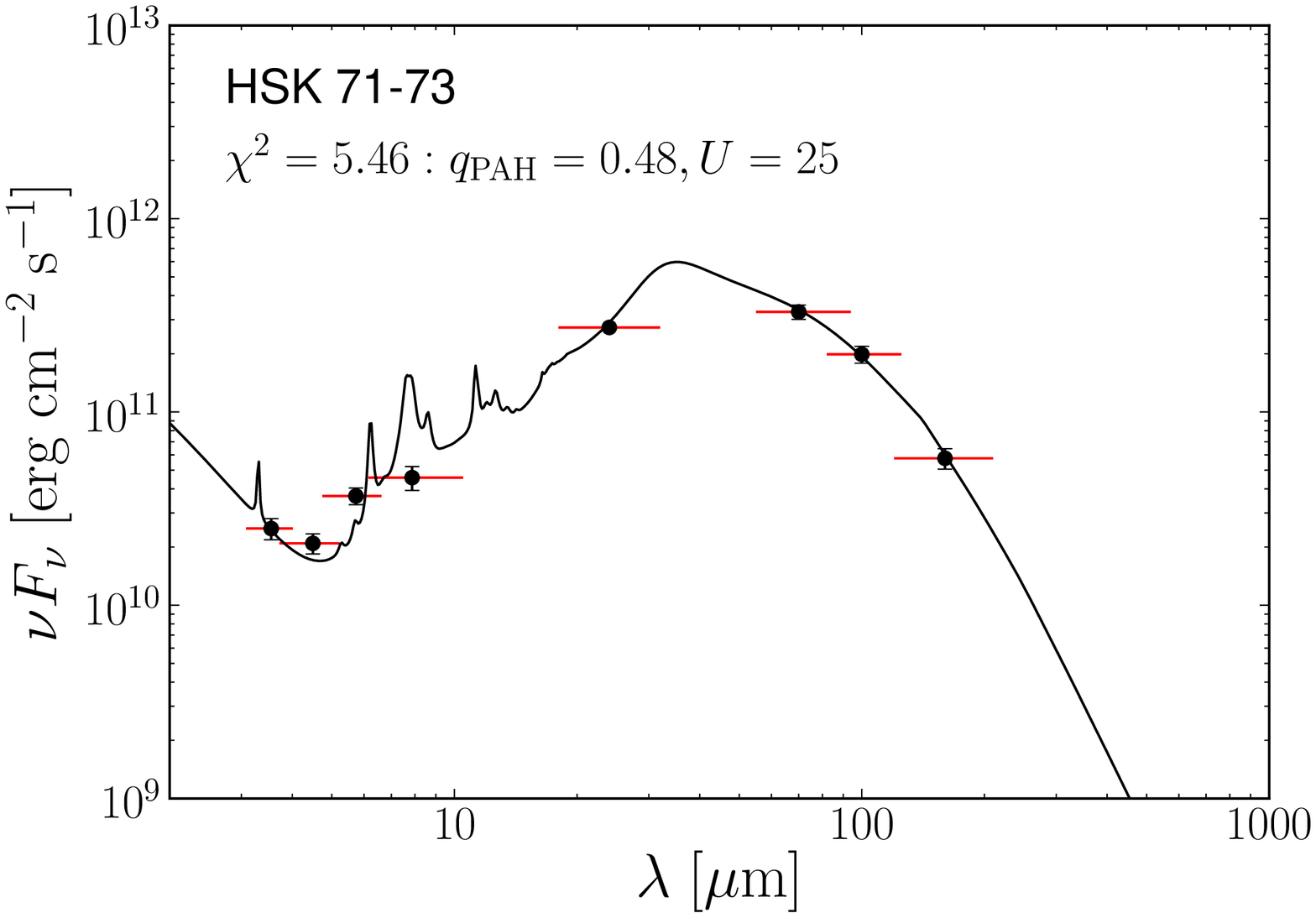}

\caption{Results of fitting the spectra of SFRs in Ho~II. \hfill}
\end{figure*}

\section*{RESULTS}

The results of fitting the spectra of 12 investigated SFRs are shown in Fig.~3,
and the corresponding parameters of the dust emission models are presented in
Table~3. As can be seen from Fig.~3, the grid of models from Draine and
Li~(2007) is insufficient to fit the spectra at a low PAH band intensity,
in particular at 8~$\mu$m. Therefore, for almost all of the SFRs, with
the exception of HSK25, HSK31, and HSK45, values of $q_{\textrm{PAH}}$ are
upper limits. Below, we will use the ratio of fluxes at~8 and 24~$\mu$m,
$F_8/F_{24}$, as a PAH abundance indicator. Sandstrom et~al.~(2010) and
Khramtsova et~al.~(2013) showed that when individual SFRs are investigated,
this ratio correlates with $q_{\textrm{PAH}}$, albeit with some scatter. Also,
we will consider the $P_8$ and $P_{24}$ indices that are the ratios
of fluxes in short-wavelength infrared bands to the cold dust flux. These
indices were introduced by Draine and Li~(2007) as near-IR emission
characteristics normalized to the total dust mass by assuming it to be
proportional to the sum of fluxes in long-wavelength bands (at 70 and
160~$\mu$m).

The brightest emission source at 8~$\mu$m in the Ho~II galaxy is HSK45, a cavity surrounded by an H$\alpha$ emission ring. It is tempting to assume
that the emission peak at 8~$\mu$m is associated with the brightest part of the
ring, but a detailed comparison of the emission map at 8~$\mu$m with the
H$\alpha$ emission map obtained with the Hubble Space Telescope shows that
the peak at 8~$\mu$m does not coincide with the ring and may be an isolated
compact background source. Since the
origin of the emission at 8~$\mu$m is unclear in this case, we excluded HSK45
from the further analysis.

\begin{figure*}[t!]
%%% Figure:4
\center{~~~~~~~~~~~~(a)~~~~~~~~~~~~~~~~~~~~~~~~~~~~~~~~~~~~~~~~~~~~~~~~~~~~~
~~~~~~~~~~~~~~~~(b)}

\includegraphics[width=\linewidth]{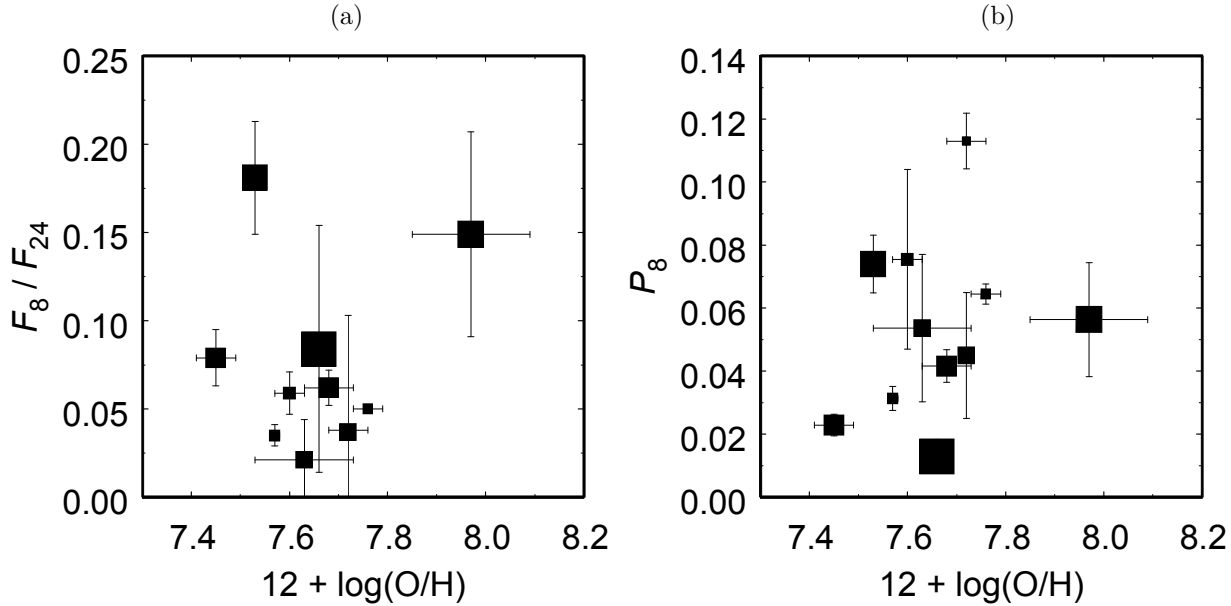}

\caption{Flux ratio $F_8/F_{24}$ (a) and $P_8$ index (b) versus metallicity.
The symbol size corresponds to the SFR age from~3.5 to 7.7~Myr. \hfill}
\end{figure*}

\begin{figure*}[t!]
%%% Figure:5
\includegraphics[width=\linewidth]{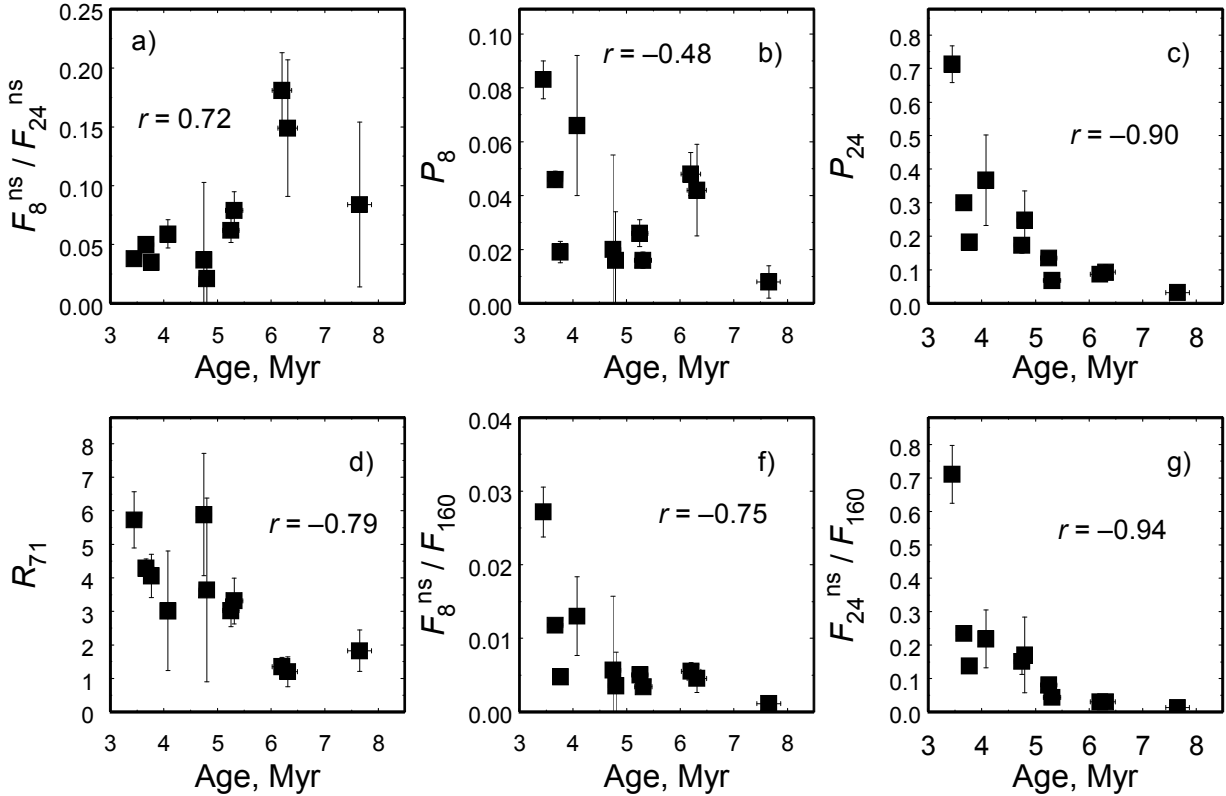}

\caption{Photometric characteristics of SFRs versus age. The Spearman rank
correlation coefficients are indicated. \hfill}
\end{figure*}

\begin{figure*}[t!]
%%% Figure:6
\includegraphics[width=\linewidth]{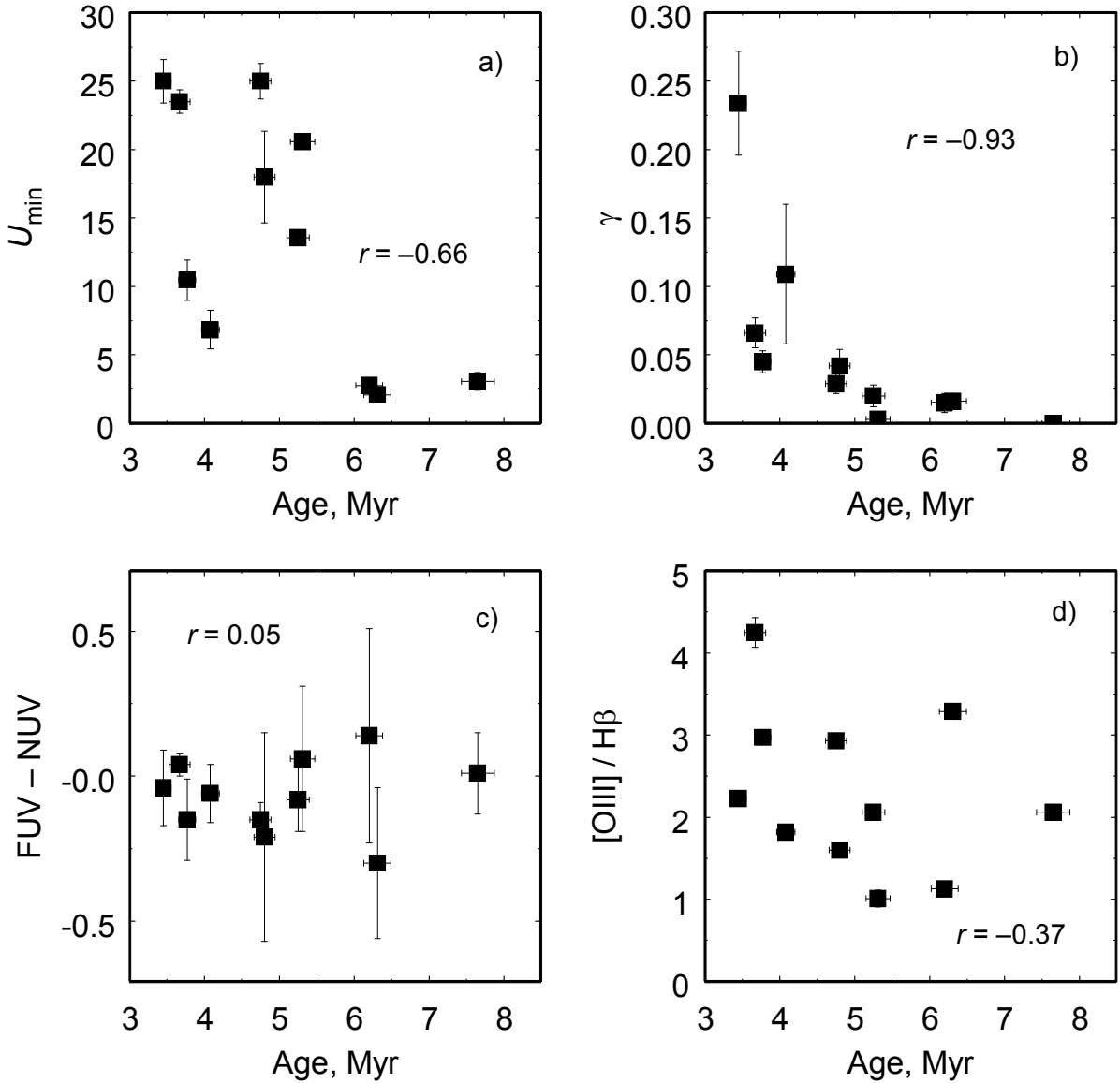}

\caption{Radiation field parameters versus age: (a) minimum
radiation field $U_{\textrm{min}}$, (b) parameter $\gamma$, (c) FUV--NUV
magnitude difference, and (d) [OIII]$\lambda$5007/H$\beta$ line ratio. The
Spearman rank correlation coefficients are indicated. \hfill}\end{figure*}

\begin{figure*}[t!]
%%% Figure:7
\includegraphics[width=\linewidth]{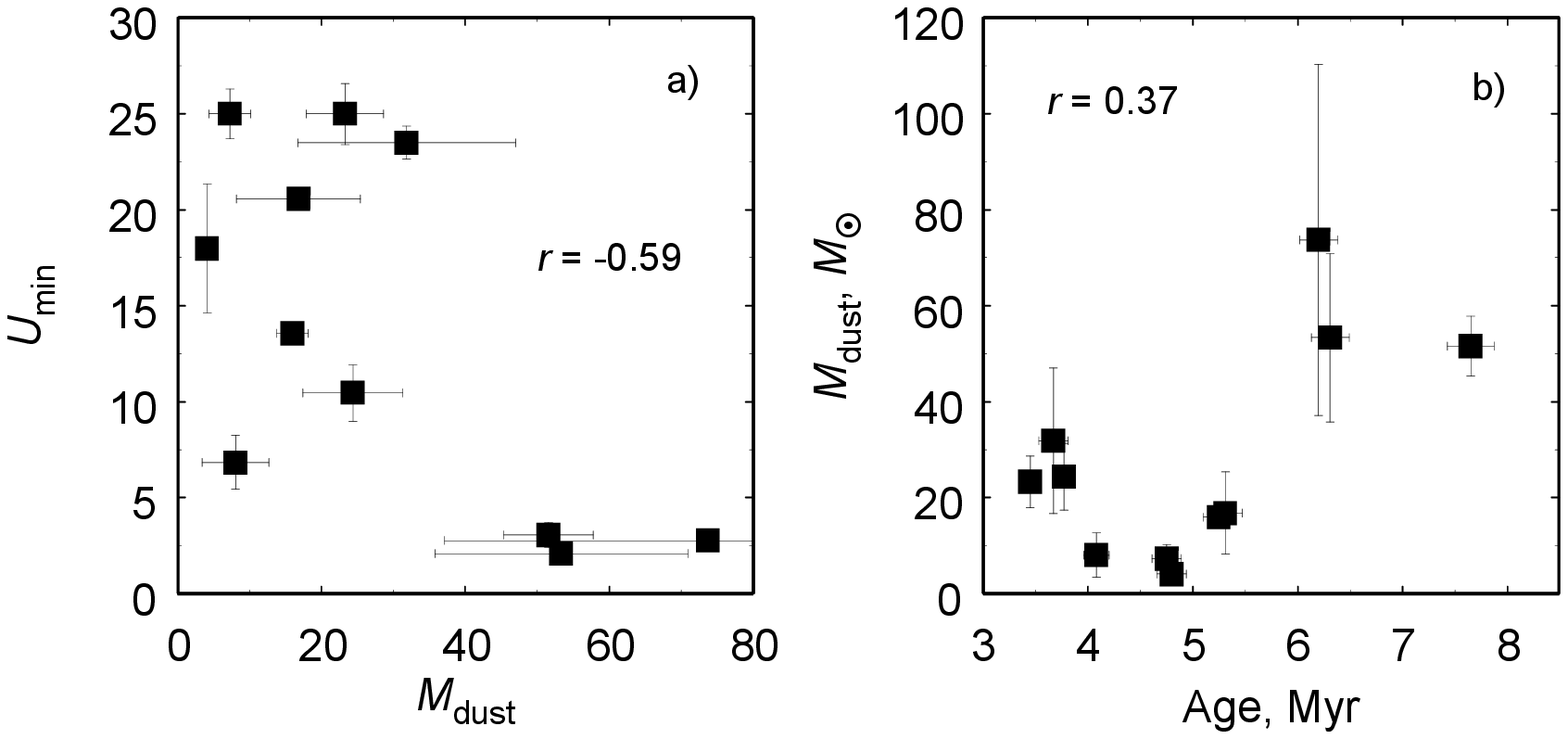}
\caption{Correlation between the dust mass, the minimum radiation
field, and the age. The Spearman rank correlation coefficients are indicated.
\hfill}
\end{figure*}

Khramtsova et~al.~(2013) pointed out that if the PAH abundance is related to
metallicity-dependent photodestruction, then it should depend not only on the
chemical composition but also on time. If SFRs are considered in a relatively
narrow range of metallicities, the PAH abundance--metallicity correlation can
be blurred by evolutionary effects. This is actually the case in
Ho~II. In Fig.~4, the flux ratio $F_8/F_{24}$ and the $P_8$ index are plotted
vs the SFR oxygen abundance; obviously, there is no correlation of both
parameters with the metallicity.

To show how the parameters characterizing $q_{\textrm{PAH}}$ change with time,
in Figs.~5a--5c we plot the $F_8/F_{24}$ flux ratio, the $P_8$ index, and the
$P_{24}$ index against the SFR age. As can be seen from the figure,
$F_8/F_{24}$, which is numerically equal to $P_8/P_{24}$, increases with age,
reflecting the growing (with time) relative contribution of particles
emitting at 8~$\mu$m. At the same time, both $P_8$ and $P_{24}$ indices
decrease with SFR age.

As has been pointed out above, the $P_8$ and $P_{24}$ indices were introduced
under the assumption that the sum of the fluxes in long-wavelength bands (at 70 and
160~$\mu$m) characterizes the total dust content. In our data, however, the
$R_{\textrm{71}}$ index characterizing the radiation field intensity depends on
age (Fig.~5d), suggesting that the flux not only at 8 and 24~$\mu$m but also in
longer-wavelength bands changes with time and, hence, cannot be an
unambiguous indicator of the dust mass. Out of the bands we consider, 
the flux at 160~$\mu$m is least dependent on age. Figures~5e and~5f show the ratios
of the fluxes at~8 and 24~$\mu$m to the flux at 160~$\mu$m. Since the emission
at 160~$\mu$m is relatively constant in the considered time interval,
these ratios hopefully reflect the evolution of the
fluxes at~8 and 24~$\mu$m more adequately. They point to a gradual weakening of the
near-IR emission as well.

Two effects can be responsible for the decreasing intensity of radiation
from PAH~particles and very small grains: the destruction of emitting particles
and the reduction in the intensity of UV~radiation heating very small grains
and exciting vibrations and bending in PAH macromolecules. The heated big grains, for which
significant destruction is not expected, must be the main source of emission at
wavelengths longer than $\sim30$~$\mu$m. Therefore, only weakening 
radiation can be responsible for the gradual decrease of the $R_{\textrm{71}}$ index.
In the model of Draine and Li~(2007), the grain-heating radiation field is
characterized by several parameters, out of which we consider the minimum SFR
radiation field, $U_{\textrm{min}}$, and the grain mass fraction $\gamma$
affected by a radiation field with intensity exceeding $U_{\textrm{min}}$. These
two parameters are plotted against time in Figs.~6a and~6b.

Both parameters decrease with time, which seems natural: as the SFR evolves,
the minimum radiation field should weaken after the initial starburst, while the
fraction of dust affected by an enhanced radiation field should decrease. The
correlation of $U_{\textrm{min}}$ with age is noticeably weaker than the
correlation of $\gamma$ with age. This is probably because $\gamma$ depends
more strongly on the number of most massive and short-lived stars. Figures~6c
and~6d show the radiation hardness parameters, the FUV--NUV magnitude
difference and the [OIII]$\lambda$~5007~\AA/H$\beta$ line ratio, for SFRs of
various ages. As can be seen, the correlation is either weak or absent in both
cases.

Figure~7a shows how the dust mass correlates with $U_{\textrm{min}}$ in the
investigated SFRs . All SFRs on this diagram can be arbitrarily divided
into two groups: with a lower mass and a stronger radiation field and with a
higher mass and a weaker radiation field. The second group includes HSK25,
HSK31, and HSK61 for which the emission peaks at about 100~$\mu$m (see Fig.~3).
For the remaining regions, the peak occurs at shorter wavelengths,
corresponding to a higher average dust temperature. Within the used model, this
implies higher dust-heating radiation intensities. In HSK25, HSK31, and HSK61,
the dust is colder than that in other regions, but their IR~flux is comparable
to that from some of the regions with ``warm'' dust. The assumption about a
higher dust mass is required for its description.

However, as has been shown above, $U_{\textrm{min}}$ correlates with the age.
Formally, this implies that the dust mass also correlates with the age in our
data (Fig.~7b). Both real factors and observational selection can be
responsible for this correlation. The absence of old low-mass regions in our
list is quite expectable: in our study we selected the SFRs based on the
brightness of emission at 8~$\mu$m, which is too weak to be detected in old
low-mass regions.

Two main factors can be responsible for the absence of young regions with a
high dust mass. First, they can actually be absent due to some peculiarities of
the current star formation episode, for example, because new starbursts were
triggered by the stimulating action of older regions. Since the ``new'' regions
are formed from the remaining matter, they may have a smaller 
scale. However, this suggestion is speculative and requires a serious study.

Second, it can also be hypothesized that this correlation is real, for example, due
to the synthesis of dust in evolved stars inside SFRs. However, the age of the
regions under consideration is too small for the ejection of dust-enriched
matter by asymptotic giant branch stars to have already begun in them. At
present, core-collapse supernovae are also considered as an important source of
dust (see, e.g., Sugerman et~al.~2006; Gomez et~al.~2012). In principle, an age
of several million years is sufficient for the evolution of the most massive
stars to have ended. However, there are no clear evidences for supernova
explosions in the kinematics of SFRs.

\section*{DISCUSSION}

The Ho~II galaxy is a promising object for investigating the PAH life cycle in
metal-poor systems. In this galaxy the emission at 8~$\mu$m (attributed to PAHs) is concentrated only in several regions coincident with
HII~complexes. Our spectroscopic observations and aperture photometry from
archival IR and UV data allow considering possible correlations between the PAH
emission and other parameters of the investigated regions.

Here, we revealed the following correlations that are important for
understanding the PAH life cycle in HII~regions:

\begin{itemize}
\item the $P_{8}$, $P_{24}$, and $R_{71}$ indices characterizing the
fractions of ``hot'' and ``warm'' dust as well as the field intensity
decrease with time;

\item the flux ratio $F_8/F_{24}$ increases with time;

\item the minimum radiation field and the fraction of dust heated by
an enhanced radiation field decrease with time;

\item the radiation field hardness parameters are virtually independent
of time.
\end{itemize}

The absence of correlation between metallicity and PAH abundance in HII~complexes
of the Ho~II galaxy implies that in this case we have an opportunity to
study age effects in a pure form. If the ratio $F_8/F_{24}$ is
considered as a measure of the relative PAH abundance (as was proposed in
several papers; see, e.g., Sandstrom et~al.~2010; Khramtsova et~al.~2013), then
we may conclude that the relative PAH abundance increases with time (Fig.~5a). Let
us consider this conclusion in more detail.

\begin{figure}[t!]
%%% Figure:8
\includegraphics[width=0.8\linewidth]{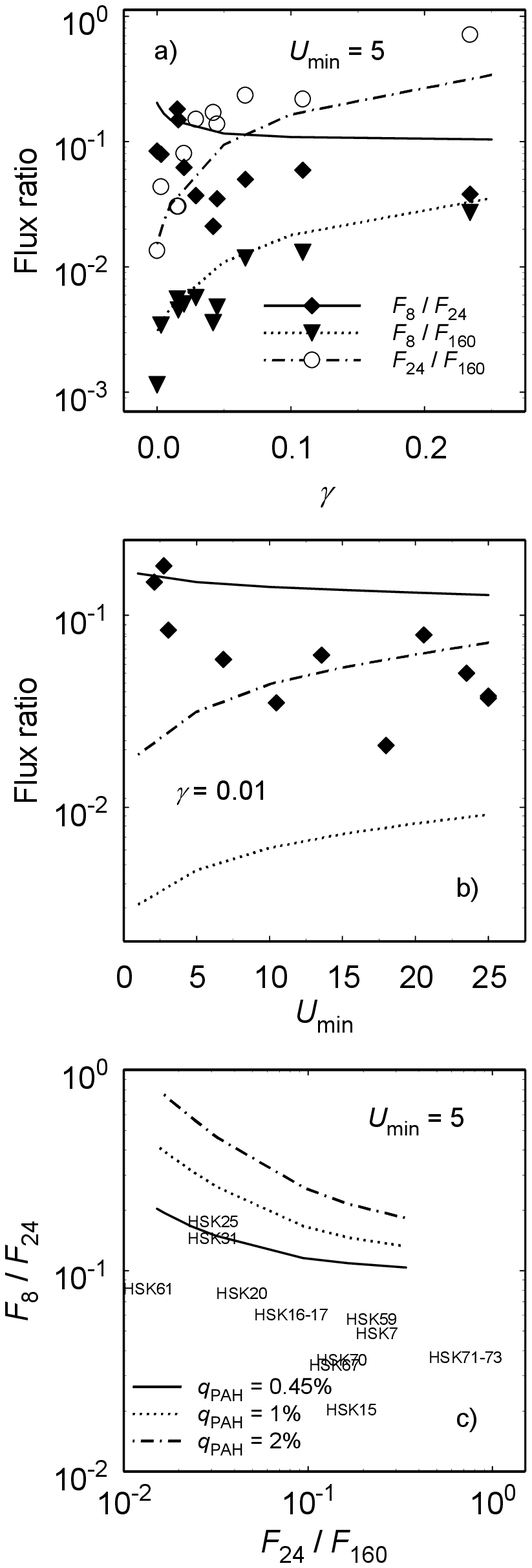}
\caption{IR~flux ratio versus parameters $\gamma$ and $U_{\textrm{min}}$
calculated using the model of Draine and Li~(2007). \hfill}
\end{figure}

The emission in almost all of the considered IR~bands weakens with time, but,
obviously, emission at 8~$\mu$m goes down more slowly than emission at 24~$\mu$m.
To unravel the reason, we calculated several spectra of SFRs with a dust mass
$M_{\textrm{dust}}=100~M_{\odot}$ and $q_{\textrm{PAH}}=0.45\%$ for
various~$\gamma$ at fixed~$U_{\textrm{min}}$ and for various~$U_{\textrm{min}}$
at fixed~$\gamma$ using the model of Draine and Li~(2007). Results of our
calculations are presented in Fig.~8 as flux ratios. Figure~8a also
shows the parameters of SFRs in the galaxy Ho~II estimated in this study.

The ratios of fluxes at~8 and 24~$\mu$m to the flux at 160~$\mu$m increase
with~$\gamma$ (Fig.~8a, the dotted and dash--dotted lines), but
$F_8/F_{160}$ depends on~$\gamma$ not so strongly as $F_{24}/F_{160}$. This
result is expectable, because the parameter~$\gamma$ in the Draine and
Li~(2007) model is mostly sensitive to the flux at 24~$\mu$m. It shows the
fraction of hot dust heated by a radiation field whose intensity exceeds the
mean background field~$U_{\textrm{min}}$, and this field is most likely associated
with the vicinities of massive stars, where the dust emission at 24~$\mu$m
should be particularly strong. The fewer such regions, the weaker emission at
24~$\mu$m is observed.

The ratio $F_8/F_{24}$ (solid line) in the model of Draine and Li~(2007)
{\emph{decreases\/}} with increasing~$\gamma$, i.e., with increasing fraction
of hot dust. However, this decrease is insignificant compared to the observed
effect. It should be noted that the observed points fall nicely on the curves
for the ratios $F_8/F_{160}$ and $F_{24}/F_{160}$, but this coincidence only
suggests that~$\gamma$ that provides the best agreement between the theoretical
and observed spectra in the used model is mainly influenced by the ratio of the
fluxes at short- and long-wavelengths and depends more weakly on
other parameters.

In Fig.~8b, the flux ratios are plotted against the
parameter~$U_{\textrm{min}}$. The same picture is observed here: as the
radiation field intensity in SFRs increases, the calculated flux ratios
$F_8/F_{160}$ and $F_{24}/F_{160}$ increase, while the ratio
$F_8/F_{24}$ decreases, but the decrease in $F_8/F_{24}$ with
increasing $U_{\textrm{min}}$ is also insignificant compared to the observed
effect in this case.

These estimates show that the ratio $F_8/F_{24}$ in ``old'' SFRs that is larger
by several times than that in ``young'' SFRs only to a small
extent can be attributed to the evolution of emission parameters ($U_{\textrm{min}}$ and
$\gamma$), in particular, to UV~field weakening.

Another factor potentially able to affect the ratio~$F_8/F_{24}$ is the destruction
of macromolecules and larger grains. The emission at 24~$\mu$m is often
associated with stochastically heated very small grains (VSGs). The more rapid
decrease of their emission relative to the PAH emission may imply that VSGs are less
stable to the destruction by UV~radiation. In addition, their destruction can itself
be a PAH source (if VSGs are PAH clusters), partially compensating for
the destruction of the macromolecules themselves.

In Fig.~8c, the $F_8/F_{24}$ ratio is plotted against $F_{24}/F_{160}$ (which correlates well with age in our
sample; see Fig.~5f) for
various~$q_{\textrm{PAH}}$. The model of Draine and Li~(2007) does not allow
calculating the emission parameters for $q_{\textrm{PAH}}<0.45\%$, but
nearly all the observed points lie below the curve for this limiting value,
approximately where the line for $q_{\textrm{PAH}}\sim0.1\%$ should have run.
Exceptions are HSK25 and HSK31, two of the three regions with an age of more than 6~Myr and
the only two regions (apart from~HSK45) for which $q_{\textrm{PAH}}$ is
$\sim1\%$, i.e., definitely above the limiting value of the model. It is this picture that
might be expected, if $q_{\textrm{PAH}}$ is initially
$\sim0.1\%$ in young SFRs of Ho~II, but then, several million years
later, increases to~$1\%$ due to the less efficient destruction of PAHs and
(or) some increase in their abundance due to VSG destruction. Thus, the ratio
$F_8/F_{24}$ and, consequently, the PAH abundance can be affected by
evolutionary effects. The scatter in the correlation between~$F_8/F_{24}$ and
metallicity may be related to the age differences.

As has been pointed out above, the regions that have not only an old age but
also the highest mass in our sample are characterized by a relatively high PAH
abundance and a less intense radiation field. The increase in the IR~flux in
these regions may be associated not so much with the pure evolution 
as with the differences in the evolution of more massive and less massive
complexes, leading to different PAH abundances. In other words, not only the
age of the complex but also its mass can be a factor affecting the PAH
abundance. However, a sample of objects that is large enough to reveal
correlations is needed to test this assumption.

HSK61, the oldest region in our sample, i.e., the region with the smallest H$\beta$
equivalent width, stands apart of the overall picture. The smallest
ratio~$F_{24}/F_{160}$ in this region agrees with the old estimated age. On the other
hand, this is the only region whose ``spectral'' age exceeds considerably the
kinematic age and is well beyond the age range determined for this region by
Stewart et~al.~(2000). Unfortunately, the available data are insufficient to
determine why HSK61 differs from other SFRs in Ho~II.

What is said above is valid, provided the emission at 24~$\mu$m is generated by
very small grains, but this interpretation is not universally accepted.
Larger grains heated to high temperatures in the vicinity of massive stars can
also be a source of this emission. Such grains are probably destroyed much more
slowly than PAHs and VSGs and the increase in~$F_8/F_{24}$ is more difficult to
explain by their destruction. An emission generation model that, in contrast to
the model of Draine and Li~(2007), would allow using an arbitrary grain
size distribution is needed to investigate this alternative. In conclusion, it
should be noted that our conclusions are based on the application of the model
by Draine and Li~(2007). At present, this model is the basic one for
determining the dust and field parameters from IR~data, but the wide
application of the model does not guarantee its universality. It may well be
that the model requires a refinement. For example, the radiation field in
HII~complexes can differ from that specified in the model. The PAH ionization
state determining the intensity of the band at 8~$\mu$m can also be different.
We plan to investigate these uncertainties in a subsequent study.

\section*{CONCLUSIONS}

Analyzing the archival IR, UV, and optical observational data for
HII~complexes in the dwarf irregular galaxy Holmberg~II, we obtained the following
results:

\begin{itemize}
\item The flux ratio $F_8/F_{24}$ can be affected by evolutionary effects related to
the destruction and formation of dust: the ratio increases as the region
evolves.

\item The emission at~8 and 24~$\mu$m gets weaker with time relative to the long-wavelength
emission and goes down with decreasing mean field intensity and
enhanced-field fraction. The emission at 24~$\mu$m decreases faster than
that at 8~$\mu$m.

\item The PAH abundance is higher in old massive complexes with a reduced radiation field.
\end{itemize}

Thus, our study shows that the fluxes at~8 and 24~$\mu$m characterizing the
emissions from PAHs and hot grains, respectively, decrease with SFR age, but
their ratio increases. This implies that the relative contribution from PAHs to
the total IR~flux increases with age. The detected increase in the ratio of the
fluxes at~8 and 24~$\mu$m is probably related to the increase in the relative
PAH fraction due to the destruction of larger grains. It should be noted that
the problem being discussed is complex and requires further comprehensive
studies in various wavelength ranges based on observations of a larger
sample of SFRs with different ages in nearby galaxies.

\section*{ACKNOWLEDGMENTS}

This work is based in part on the observational data from the BTA (SAO RAS)
telescope and those taken from the archives of the Spitzer, Herschel, and GALEX
space telescopes. This study was financially supported by the Russian
Foundation for Basic Research (project nos.~12-02-31356 mol\_a, 12-02-31452
mol\_a and 14-02-00604).

\section*{\centerline{REFERENCES}}

\begin{enumerate}

\item G.~Aniano, B.~T.~Draine, K.~D.~Gordon, and K.~Sandstrom, Publ. Astron. Soc. Pacif. \textbf{123}, 1218 (2011)

\item I.~Bagetakos, E.~Brinks, F.~Walter, et~al., Astron. J.~\textbf{141}, 23 (2011).

\item D.~Calzetti, EAS Publ. Ser. \textbf{46}, 133 (2011).

\item M.~V.~F.~Copetti, M.~G.~Pastoriza, and H.~A.~Dottori, Astron. Astrophys. \textbf{156}, 111 (1986).

\item B.~T.~Draine and A.~Li, Astrophys. J.~\textbf{657}, 810 (2007).

\item B.~T.~Draine, D.~A.~Dale, G.~Bendo, K.~D.~Gordon, et al., Astrophys. J.~\textbf{663}, 866 (2007).

\item O.~V.~Egorov, T.~A.~Lozinskaya, and A.~V.~Moiseev, Mon. Not. R.~Astron. Soc. \textbf{429}, 1450 (2013).

\item C.~W.~Engelbracht, K.~D.~Gordon, G.~H.~Rieke, et al., Astrophys. J.~\textbf{628}, 29 (2005).

\item H.~L.~Gomez, O.~Krause, M.~J.~Barlow, B.~M.~Swinyard, et al., Astrophys. J.~\textbf{760}, 96 (2012).

\item K.~D.~Gordon, Ch.~W.~Engelbracht, G.~H.~Rieke, et al., Astrophys. J.~\textbf{682}, 336 (2008).

\item P.~Hodge, N.~V.~Strobel, and R.~C.~Kennicutt, Publ. Astron. Soc. Pacif. \textbf{106}, 309 (1994).

\item L.~K.~Hunt, T.~X.~Thuan, Y.~I.~Izotov, and M.~Sauvage, Astrophys. J.~\textbf{712}, 164 (2010).

\item D.~A.~Hunter, F.~C.~Gillett, J.~S.~Gallagher, et al., Astrophys. J.~\textbf{303}, 171 (1986).

\item R.~C.~Kennicutt, L.~Armus, G.~Bendo, D.~Calzetti, et al., Publ. Astron. Soc. Pacif. \textbf{115}, 928 (2003).

\item R.~C.~Kennicutt, D.~Calzetti, D.~Aniano, et al., Publ. Astron. Soc. Pacif. \textbf{123}, 1347 (2011).

\item M.~S.~Khramtsova, D.~S.~Wiebe, P.~A.~Boley, and Ya.~N.~Pavlyuchenkov, Mon. Not. R.~Astron. Soc. \textbf{431}, 2006 (2013).

\item C.~Leitherer, D.~Schaerer, J.~D.~Goldader, et al., Astrophys. J.~\textbf{123}, 3 (1999).

\item A.~K.~Leroy, F.~Walter, F.~Bigiel, et al., Astron. J.~\textbf{137}, 4670 (2009).

\item E.~M.~Levesque, L.~J.~Kewley, and K.~L.~Larson, Astron. J.~\textbf{139}, 712 (2010).

\item M.-M.~Mac~Low and R.~McCray, Astrophys. J. \textbf{324}, 776 (1988).

\item S.~C.~Madden, F.~Galliano, A.~P.~Jones, and M.~Sauvage, Astron. Astrophys. \textbf{446}, 877 (2006).

\item A.~R.~Marble, C.~W.~Engelbracht, L.~van~Zee, et al., Astrophys. J.~\textbf{715}, 506 (2010).

\item J.~S.~Mathis, P.~G.~Mezger, and N.~Panagia, Astron. Astrophys. \textbf{128}, 212 (1983).

\item J.~Moustakas, R.~C.~Kennicutt, C.~A.~Tremonti, et al., Astrophys. J.~Suppl. Ser. \textbf{190}, 233 (2010).

\item D.~Puche, D.~Westpfahl, E.~Brinks, and J.-R.~Roy, Astron. J.~\textbf{103}, 1841 (1992).

\item
 K.~M.~Sandstrom, A.~D.~Bolatto, B.~T.~Draine, et al., Astrophys. J.~\textbf{715}, 701 (2010).
 
\item D.~Schaerer and W.~D.~Vacca, Astrophys. J. \textbf{497}, 618 (1998).

\item G.~Stasi\'nska and C.~Leitherer, Astrophys. J.~Suppl. Ser. \textbf{107}, 661 (1996).

\item S.~G.~Stewart, M.~N.~Fanelli, G.~G.~Byrd, et al., Astrophys. J.~\textbf{529}, 201 (2000).

\item B.~E.~K.~Sugerman, B.~Ercolano, M.~J.~Barlow, et al., Science \textbf{313}, 196 (2006).

\item F.~Walter, E.~Brinks, W.~J.~G.~de~Blok, et al., Astron. J.~\textbf{136}, 2563 (2008).

\item J.~C.~Weingartner and B.~T.~Draine, Astrophys. J.~\textbf{548}, 296 (2001).

\item D.~R.~Weisz, E.~D.~Skillman, J.~M.~Cannon, et al., Astrophys. J.~\textbf{689}, 160 (2008).

\item D.~R.~Weisz, E.~D.~Skillman, J.~M.~Cannon, et al., Astrophys. J.~\textbf{704}, 1538 (2009).

\item D.~S.~Wiebe, O.~V.~Egorov, and T.~A.~Lozinskaya, Astron. Rep. \textbf{55}, 585 (2011).

\end{enumerate}

\textit{Translated by V.~Astakhov}

\end{document}